\documentclass[12pt]{article}
\pdfoutput=1
\usepackage{pstricks}
\usepackage{color}
\usepackage{amssymb,amsmath,bm,bbm}
\usepackage{epsf}
\usepackage{epsfig}
\usepackage{afterpage}
\usepackage{longtable}
\usepackage{cite}
\usepackage{latexsym,mathrsfs,dsfont}
\usepackage{graphics}
\usepackage{url}
\usepackage{paralist}
\usepackage{bbold}
\usepackage[colorlinks=true, urlcolor=blue]{hyperref}
\usepackage{appendix}
\newcommand{\nn}{\nonumber}

\newcommand{\spur}[1]{\not\! #1 \,}
\newcommand{\be}{\begin{equation}}
\newcommand{\ee}{\end{equation}}
\newcommand{\bea}{\begin{eqnarray}}
\newcommand{\eea}{\end{eqnarray}}

\newcommand{\dd}{\displaystyle}
\numberwithin{equation}{section}

\begin{document}

\begin{flushright}
    {BARI-TH/23-746}
\end{flushright}

\medskip

\begin{center}
{\Large\bf
  \boldmath{On the decay mode $\Lambda_b \to X_s \gamma$}}
\\[0.8 cm]
{\large P.~Colangelo$^a$, F.~De~Fazio$^a$ and F.~Loparco$^a$
 \\[0.5 cm]}
{\small
$^a$
Istituto Nazionale di Fisica Nucleare, Sezione di Bari, Via Orabona 4,
I-70126 Bari, Italy
}
\end{center}

\vskip0.5cm


\begin{abstract}
\noindent
We study the inclusive  $H_b \to X_s \gamma$ decay with $H_b$  a  beauty baryon,   in particular   $\Lambda_b$,  employing an expansion in the 
 heavy quark mass at ${\cal O}(m_b^{-3})$ at leading order in $\alpha_s$, keeping the dependence on the hadron spin.  For a polarized baryon we compute the distribution $\displaystyle\frac{d^2\Gamma}{dy \, d \cos \theta_P}$, with  $y=2E_\gamma/m_b$, $E_\gamma$  the photon energy and $\theta_P$ the angle between the  baryon spin vector and the photon momentum in the $H_b$ rest-frame. We discuss the correlation between the baryon  and  photon polarization,  and show that  effects of physics beyond the Standard Model can modify the photon polarization asymmetry.    
We  also discuss a method to treat the singular terms in the photon energy spectrum obtained by the OPE.

\end{abstract}

\thispagestyle{empty}
\newpage
\section{Introduction}

The processes induced at the quark level by the $b \to s \gamma$ transition are recognized as a powerful testground of the  Standard Model (SM)  \cite{Shifman:1976de,Bertolini:1986th,Grinstein:1987vj}. They occur at loop-level in SM and are sensitive to  heavy particle exchanges.   Upon  integration of the heavy quanta, an effective Hamiltonian   is obtained in terms of local operators and Wilson coefficients \cite{Cella:1994px,Misiak:2015xwa}. Physics beyond the Standard Model induces new operators  with respect  to the  SM ones and modifies the Wilson coefficients, hence measurements of various observables   tightly  constrain  operators  and  coefficients. This allows to probe the  SM and select the possible extensions\cite{Grinstein:1987pu,Hou:1987kf,Gabbiani:1996hi,Everett:2001yy,Borzumati:2003rr,Buras:2003mk,Blanke:2006sb,Misiak:2017bgg}. 

The radiative $b \to s$ transition  has been intensively analyzed in theory.\footnote{A detailed discussion and a list of  references can be found in \cite{Buras:2020xsm}.} Strong experimental efforts have  also been devoted since the first observation  of the  $B \to K^*(892)\gamma$ mode  \cite{CLEO:1993nic}.  Several measurements of exclusive processes are now available, namely ${\cal B}(B^+ \to K^{*+}(892) \gamma)$ and ${\cal B}(B^0 \to K^{*0}(892) \gamma)$, and the rates of $B \to (K_1(1400),K_2^*(1430),K_3^*(1780)) \gamma$,  
 $B \to K \eta \gamma$, $B_s \to \phi(1020) \gamma$ \cite{HFLAV:2022pwe,ParticleDataGroup:2022pth}. 
Time-dependent CP asymmetries  for decaying neutral  mesons have been investigated \cite{HFLAV:2022pwe}. 
 For baryons,   the rate and the photon polarization  of $\Lambda_b \to \Lambda \gamma$ have been measured   
\cite{LHCb:2019wwi,LHCb:2021byf},   and  an upper bound  has been put to  ${\cal B}(\Xi_b^-\to\Xi^-\gamma)$  \cite{LHCb:2021hfz}.  For such exclusive processes, the hadronic uncertainties are related to the form factors parametrizing  the matrix elements of the local operators at $q^2=0$, with $q$  the photon momentum \cite{Colangelo:1993ux}.
 
Among the inclusive $H_b \to X_s \gamma$ modes  the prime example is  $\bar B \to X_s \gamma$. Here we  focus on $H_b$ a  baryon, in particular on $\Lambda_b \to X_s \gamma$. The peculiarity of  the inclusive modes consists in the possibility, invoking quark-hadron duality, of  exploiting a well defined 
 theoretical framework based on controlled expansions in QCD quantities $\alpha_s(m_b)$ and $1/m_b$ (the heavy quark expansion, HQE) to compute  decay rates,  decay distributions and moments of the distributions.  At the leading order in the heavy quark expansion the partonic result is recovered,  NLO terms involve  nonperturbative  corrections. 
In particular, the combination of Operator Product Expansion (OPE) and  Heavy Quark Effective Theory (HQET) \cite{Neubert:1993mb} allows to express  the inclusive
decay widths as an expansion in the inverse heavy quark mass. Based on this approach,   inclusive  semileptonic modes  of hadrons comprising a single heavy quark  are exploited to access  fundamental  parameters,  the heavy quark masses and elements of the Cabibbo-Kobayashi-Maskawa (CKM)  mixing matrix.  Input quantities are the hadronic matrix elements of  local operators, starting from   the kinetic energy and  the chromomagnetic operators,  which are  defined in the following.  Differential distributions  together with  other  observables can also be described. In  $ B \to X_s \gamma$, 
direct CP asymmetries sensitive to new physics (NP) effects can be studied upon accounting for  long-distance contributions  \cite{Kagan:1998bh,Benzke:2010tq}. 

There are several issues in inclusive modes induced by the $b \to s \gamma$ transition needing to be considered.
The actual expansion parameter is the inverse of the energy released in the  process. Such energy is  ${\cal O}(m_b$)
in a portion of the phase-space, but  in some regions its inverse  is no longer small.  Signals about the reliability of the method show up as  singularities in differential distributions.
This occurs in the calculation of perturbative corrections
to the spectra, where Sudakov terms appear.\footnote{Sudakov terms (referred to as  Sudakov shoulders) located in the middle of the phase-space can also be present when singularities  from real and virtual perturbative corrections do not compensate each other \cite{Catani:1997xc,DeFazio:1999ptt}.}  Singular terms appear at higher orders in the HQE in form of the delta distribution and its derivatives, the argument of which vanishes in the regions corresponding  to the endpoints of the differential distributions determined in the partonic kinematics, different from the hadronic  kinematics. 
The gap between the two borders is governed by nonperturbative physics responsible of  bound state effects. 
Such effects can be related
to the Fermi motion of the heavy quark in the decaying hadron, and can be accounted for introducing a shape function which encodes information on the distribution of the $b$
quark  residual momentum in   the hadron \cite{Neubert:1993ch,Neubert:1993um,Bigi:1993ex}. The same
function enters in the description of  different inclusive modes, and  it affects the photon energy
spectrum in $H_b \to X_s \gamma$. In case of $B$,  the moments of the shape function have been constrained using  measurements \cite{HFLAV:2022pwe,Bernlochner:2020jlt}.

 Another source of uncertainty in inclusive $b \to s \gamma$ processes are the  { resolved photon} contributions  \cite{Benzke:2010js},  related to the  photon couplings  different from  the effective weak interaction vertex.  The most important operators giving rise to these contributions are $O_2$ and $O_8$ (in the notation specified in the following Section). Such effects appear  at ${\cal O}(1/m_b)$ and  produce  contributions to the total decay width which  are not  described  by the HQE  \cite{Donoghue:1995na,Lee:2006wn}. 
They can be   expressed in terms of subleading shape functions   \cite{Benzke:2010js}. The resolved photon contributions include a nonperturbative term proportional to the matrix element $\mu_G^2$ of the chromomagnetic operator,  related to the  gluon-photon  penguin mechanism,
for which the short distance scale is $1/m_c$ rather than $1/m_b$  \cite{Voloshin:1996gw,Ligeti:1997tc,Grant:1997ec,Buchalla:1997ky}. This is 
 estimated to be  small in $\bar B \to X_s \gamma$. For  $\Lambda_b$ the matrix element of the chromomagnetic operator vanishes, therefore  the comparison of the inclusive   $\Lambda_b$ and $B$ radiative decay widths provides a way to shed light on the role of such  corrections. 

In radiative  modes the photon spectrum can be measured above an energy threshold. For $\bar B$ the
HFLAV  Collaboration quotes ${\cal B}({\bar B} \to X_s \gamma)=(3.49 \pm 0.19)\times 10^{-4}$ for $E_\gamma>1.6 $ GeV, with both charged and neutral  mesons included in the average. The  SM result  is ${\cal B}({\bar B} \to X_s \gamma)=(3.36 \pm 0.23) \times 10^{-4}$ for the same  threshold \cite{Misiak:2015xwa}. A global fit including all data on the photon energy spectrum in ${\bar B} \to X_s \gamma$ has been performed by the SIMBA Collaboration \cite{Bernlochner:2020jlt}.
 
The interest for inclusive $\Lambda_b \to X_s \gamma$  relies on the possibility in baryon modes  to investigate observables sensitive to the spin of the decaying hadron. 
 This is important for the planned new lepton facilities, since heavy baryons with a b-quark  produced  from $Z^0$ and top-quark  decays are expected to have a sizable polarization, as observed at LEP
\cite{BUSKULIC1996437,Abbiendi:1998uz,2000205}.
The application of  HQE to baryons requires new information, namely the operator matrix elements for specified hadron spin. Such matrix elements have been  analyzed  in \cite{Colangelo:2020vhu}.
Moreover, the leading and subleading shape functions are different  for different hadrons, namely for
 $\Lambda_b$  and  $B$, and require dedicated considerations. 
 
 In the present study we focus on two issues. The first one is the dependence on the heavy baryon spin in a double differential decay distribution, considering hadronic matrix elements at ${\cal O}(\/1/m_b^3)$, for the leading operator in the SM effective weak Hamiltonian and for a single new physics (NP) operator, studying the correlations between the baryon and photon polarization. The second one is  a way  to treat the singular terms in the inclusive photon spectrum to reconstruct the  $\Lambda_b$  leading shape function, a method which can be systematically applied when higher order terms in the heavy quark expansion are computed. 

The plan of the paper is the following. Section~\ref{sec:Hamil}  includes the $b \to s \gamma$ low-energy Hamiltonian  with  SM operators and operators  obtained  in extensions of the Standard Model.  In Sec.~\ref{OPE} we  describe the  application of the  HQE to the inclusive $H_b \to X_s \gamma$ process with $H_b$ a baryon, in particular $\Lambda_b$, keeping the dependence on the baryon spin. 
In Sec.~\ref{sec:polarization} we investigate the correlation between the  photon and  $\Lambda_b$ polarizations. A treatment
 of the singular terms  in the double differential decay rate is discussed in Sect.~\ref{sf}.  Details are collected in the Appendices. In the  last Section we present our conclusions and the perspectives for further progress.

\section{$b \to s \gamma$ effective Hamiltonian}\label{sec:Hamil}
The  low-energy Hamiltonian governing the $\Delta B =-1$, $\Delta S = 1$  $b \to s \gamma$  transition can be written as
\bea
H_{\rm eff}^{b \to s \gamma}=-\,4\,\frac{G_F }{\sqrt{2}} V_{tb} V_{ts}^* \, && \sum_{i}\left[C_i(\mu) O_i  +C_i^{\prime}(\mu) O_i^{\prime  } \right]\, , \,\,\,\,\,\, \label{hamil}
\eea
with $i=1, \dots 8$ and $i=15,\dots 20$.
 $G_F$  is the Fermi constant and $V_{jk}$ are elements of the CKM  matrix.  Doubly Cabibbo-suppressed terms proportional to $V_{ub} V_{us}^*$ have been neglected in ~(\ref{hamil}). 
 The effective Hamiltonian  comprises the magnetic penguin operators
\bea
O_7&=&\frac{e}{16 \pi^2} [{\bar s}\sigma^{\mu \nu}(m_s P_L + m_b P_R)\,b] F_{\mu \nu} \, , \label{O7} \\
O_8&=&\frac{g_s}{16 \pi^2}\Big[{\bar s}_{ \alpha} \sigma^{\mu \nu} \Big({\lambda^a \over 2}\Big)_{\alpha \beta} (m_s P_L + m_b P_R)  b_{ \beta}\Big] 
      G^a_{\mu \nu}  \, ,  \label{O8} 
      \eea
with  $P_{R,L}=\displaystyle\frac{1 \pm \gamma_5}{2}$  helicity projectors,  $\alpha,\beta$  colour indices,
 $\lambda^a$  the Gell-Mann matrices.  $F_{\mu \nu}$ and $G^a_{\mu \nu}$  are the
electromagnetic and  gluonic field strengths,  $e$ and $g_s$ the
electromagnetic and strong coupling constants, $m_b$ and $m_s$  the $b$ and $s$ quark mass. 
 The Hamiltonian also comprises the current-current operators  $O_{1,2}$,
\bea
O_1 &=& ({\bar s}_\alpha \gamma^\mu P_L \, c_\beta)({\bar c}_\beta \gamma_\mu P_L \, b_\alpha) \, , \label{O1} \\
O_2 &=& ({\bar s} \gamma^\mu P_L \, c)({\bar c} \gamma_\mu P_L \, b) , \label{O2}
\eea
and the  QCD penguin operators $O_{i=3,\dots 6}$, 
\bea
O_3&=&({\bar s} \gamma^\mu P_L \, b) \sum_q ({\bar q} \gamma^\mu P_L \, q) \, , \hskip 1 cm 
O_4=({\bar s}_\alpha \gamma^\mu P_L \, b_\beta) \sum_q ({\bar q}_\beta \gamma^\mu P_L \, q_\alpha) \, , \label{O34} \\
O_5&=&({\bar s} \gamma^\mu P_L \, b) \sum_q ({\bar q} \gamma^\mu P_R \, q) \, , \hskip 1 cm 
O_6=({\bar s}_\alpha \gamma^\mu P_L \, b_\beta) \sum_q ({\bar q}_\beta \gamma^\mu P_R \, q_\alpha) . \label{O56}
\eea
 The sum in (\ref{O34})-(\ref{O56}) runs over the flavours $q=u,d,s,c,b$.
The remaining operators,  absent in SM,  are analogous to the QCD penguins but have a scalar or tensor structure
\cite{Borzumati:1999qt}:
 \begin{align}
  O_{15}^q &=  (\bar{s}P_R b) \sum_q(\bar{q} P_R q)\,, &
O_{16}^q &=  (\bar{s}_\alpha P_R b_\beta) \sum_q(\bar{q}_\beta P_R q_\alpha)\,, \notag \\
O_{17}^q &=  (\bar{s}P_R b) \sum_q(\bar{q} P_L q)\,, &
  O_{18}^q &=  (\bar{s}_\alpha P_R b_\beta)\sum_q (\bar{q}_\beta P_L q_\alpha)\,, \\
  O_{19}^q &=  (\bar{s} \sigma^{\mu\nu} P_R b) \sum_q(\bar{q} \sigma_{\mu\nu} P_R q)\,, &
 O_{20}^q &=  (\bar{s}_\alpha \sigma^{\mu\nu} P_R b_\beta) \sum_q(\bar{q}_\beta \sigma_{\mu\nu} P_R q_\alpha)\, . \notag
\end{align}
The primed operators  have opposite chirality with respect to the unprimed ones. 

In  SM the process $b \to s \gamma$ is described by  photon penguin diagrams, with the photon coupled either to the intermediate  fermion  or to the $W^\pm$,   giving rise to the magnetic operator $O_7$. This is the only operator contributing at lowest order in QCD.
The renormalization group evolution to the scale $\mu_b \simeq {\cal O}(m_b)$ also involves the magnetic gluon penguin operator $O_8$ and the operators $O_{1,\dots 6}$. Their mixing into $O_7$ generates large logarithms  producing a strong enhancement of the rate. The anomalous dimension matrix  governing the mixing turns out to be regularization scheme dependent. One can get rid of such a dependence   defining an effective coefficient
 $C_7^{\rm eff}(\mu_b)$ which includes contributions of $O_{1,\dots 6}$ \cite{Buras:1993xp}. In this way  $O_7$ turns out to be  the dominant contribution to $b \to s \gamma$, with the  SM Wilson coefficients known at NNLO in QCD \cite{Misiak:2018cec,Buras:2020xsm}.
Extensions of the SM can also induce the operators  $O_{15}^q-O_{20}^q$ and the primed operators  in the low-energy Hamiltonian.

In this paper  we work at the leading order in $\alpha_s$ so that the only operator mediating the  $b \to s \gamma$ transition is $O_7$ in  SM and possibly  $O_7^{\prime}$ beyond  SM, the effect of $O_{1,\dots 6}^{(\prime)}$ being included in the effective coefficients. Therefore, we consider the effective Hamiltonian at the scale $\mu_b$ consisting  of only two operators,
\begin{equation}
H_{\rm eff}^{b \to s \gamma}=-\,4\,{G_F \over \sqrt{2}} V_{tb} V_{ts}^* \, \Big\{C_7^{\rm eff}O_7+C_7^{\prime \rm eff} O_7^\prime \Big\} . \label{Hbsg}
\end{equation}
We  do not consider operators from  NLO electroweak corrections.

$H_{\rm eff}^{b \to s \gamma}$ can be recast in a  way suitable for  the heavy quark expansion:
\begin{equation}
H_{\rm eff}^{b \to s \gamma}=-\,4\,{G_F \over \sqrt{2}} \lambda_t \, \frac{e}{16 \pi^2} \,\sum_{i=7,7^\prime} C_i^{\rm eff} \, J^{i}_{\mu \nu}\, F^{\mu \nu}  \,\, , \label{hnew}
\end{equation}
where $\lambda_t=V_{tb} V_{ts}^*$, $J^{i}_{\mu \nu}=[{\bar s}\sigma_{\mu \nu}(m_s (1 - P_i) + m_b P_i)\,b]$ and $P_i=P_R$ for $i=7$, $P_i=P_L$ for $i=7^\prime$.
In the next Section  we compute the inclusive  width $\Gamma (H_b \to X_s \gamma)$ using the Hamiltonian  (\ref{hnew}), as done for the semileptonic modes in \cite{Colangelo:2020vhu}.

\section{Inclusive decay width}\label{OPE}
To  describe the inclusive mode $H_b(p,s) \to X_s(p_X) \gamma (q,\epsilon)$ 
we preliminarly define 
\be
{\cal F}^{MN}=
{\cal F}^{\mu \nu \mu^\prime \nu^\prime} =\sum_\epsilon 4q^\nu\, q^{\nu^\prime}\,\epsilon^\mu \epsilon^{*\mu^\prime}=
-4q^\nu\, q^{\nu^\prime}\, g^{\mu \mu^\prime} ,
\label{em}
\ee
where $q$ and $\epsilon$ are the photon momentum and polarization four-vector, respectively,  using the compact notation $M={\mu \, \nu},\, N={\mu^\prime \, \nu^\prime}$.
To obtain the results specifying the photon polarization we also define
\bea
{\cal F}^{MN}_+&=&
4q^\nu\, q^{\nu^\prime}\,\epsilon_+^\mu \epsilon_+^{*\mu^\prime} \,\, ,\nn \\
{\cal F}^{MN}_-&=
&4q^\nu\, q^{\nu^\prime}\,\epsilon_-^\mu \epsilon_-^{*\mu^\prime} \,\, , 
\label{empm}
\eea
where 
\begin{equation}
\epsilon_\pm = \mp \, \frac{1}{\sqrt{2}} \, ( 0, 1, \pm i, 0 ) \,\, .
\end{equation}
The differential inclusive decay width can be written as
\be
d\Gamma= [dq] \,  \frac{G_F^2 |\lambda_t|^2}{8m_{H_b}} \frac{\alpha}{\pi^2} \,\sum_{i,j=7,7^\prime} C_i^{\rm eff * } C_j^{\rm eff} \,\, W^{ij}_{MN} {\cal F}^{MN} , \label{dgamma}
\ee
with   $[dq]=\displaystyle\frac{d^3 q}{(2\pi)^3 2q^0}$. 
 By the optical theorem,  the hadronic tensor $W^{ij}_{MN}$ is  related to
 the discontinuity of the forward   scattering amplitude
\bea
T^{ij}_{MN}&=&i\,\int d^4x \, e^{-i\,q \cdot x} \langle H_b(p,s)|T[ J^{i\dagger}_M (x) \,J^{j}_N (0)] |H_b(p,s) \rangle\,\,\label{Tij-gen}
\eea
 across the cut corresponding to the process $H_b(p,s) \to X_s(p_X) \gamma (q,\epsilon)$:
\be
W^{ij}_{MN}=\frac{1}{\pi}{\rm Im} \, T^{ij}_{MN} .
\ee 
The range of the invariant mass $p_X^2$  of the states produced in $B$ and $\Lambda_b$ decays (with $p_X=p-q$) is $p_X^2 \in [m^2_{K^*},\, m_B^2]$ and $p_X^2 \in [m_\Lambda^2,\,m_{\Lambda_b}^2]$, respectively.  For $m_b \to \infty$, $p_X^2$ is  almost always large enough to exploit the short distance limit $x \to 0$ in Eq.~(\ref{Tij-gen}), thus allowing a computation of  
$T^{ij}$ and $W^{ij}$ by an OPE  with  expansion parameter  $ \dd \frac{1}{m_b}$   \cite{Chay:1990da,Bigi:1993fe}. The first term of the expansion describes  the free beauty quark decay, the partonic result. The expansion is valid in the largest part of the  phase-space, it fails in the region with small $p_X^2$, therefore  a reliable computation of the decay width and of moments of the decay distributions can be carried out. For spectra, the result obtained by the short distance OPE needs to be smeared:  in Section \ref{sf} we 
discuss a way to implement the smearing.

The procedure for the derivation of the OPE for Eq.\eqref{Tij-gen} is summarized in Appendix \ref{newappOPE}. 
Using the definition 
\be
{\tilde {\cal T}}^{ij}=T^{ij}_{MN}{\cal F}^{MN} \,\, 
\ee
we obtain the expression:
\bea
\sum_{i,j=7,7^\prime} C_i^{\rm eff *} \, C_j^{\rm eff} \, {\tilde {\cal T}}^{ij} &=& \bigg[ ( m_b^2 + m_s^2 ) \, ( | C_7^{\rm eff} |^2 + | C_7^{\prime \,\rm eff} |^2 ) + 4 \, m_b \, m_s \, {\rm Re}[C_7^{\rm eff} \, C_7^{\prime \,\rm eff *} ] \bigg] \, \tilde{T} \quad \nn \\
&+& ( m_b^2 - m_s^2 ) \, ( | C_7^{\rm eff} |^2 - | C_7^{\prime \,\rm eff} |^2 ) \, \tilde{S}  ,
\eea
where
\bea
{\tilde T}&=&16 \, m_H \, (v \cdot q )^2 \, \sum_{n=1}^4 \left(\frac{m_b}{\Delta_0} \right)^n \,  {\tilde T}_n  , \nn \\
{\tilde S}&=&16 \, m_H \, (v \cdot q ) \, (q \cdot s \, )\sum_{n=1}^4 \left(\frac{m_b}{\Delta_0} \right)^n \,  {\tilde S}_n  ,\label{Ttilde}
\eea
with $v$ and $\Delta_0$ defined in Appendix  \ref{newappOPE}.
The four terms in \eqref{Ttilde} involve the hadronic parameters ${\hat \mu}_\pi^2$,  ${\hat \mu}_G^2$,   ${\hat \rho}_D^3$  and  ${\hat \rho}_{LS}^3 $ defined in the same Appendix, and read:
\bea
{\tilde T}_1 &=&
1 + \frac{5}{6 \, m_b^2} \big[  {\hat \mu}_\pi^2  - {\hat \mu}_G^2 \big]-\frac{2}{3m_b^3} \big[ {\hat \rho}_D^3+{\hat \rho}_{LS}^3 \big]  ,
\\
{\tilde S}_1 &=&
1+\frac{1}{4 m_b^2} \big[  {\hat \mu}_\pi^2  +{\hat \mu}_G^2 \big]+\frac{1}{6m_b^3} \,{\hat \rho}_D^3  ,
\\
{\tilde T}_2 &=&\frac{7 v \cdot q}{3m_b^2}{\hat \mu}_\pi^2+
\frac{1}{3 m_b^2} (4m_b-5v \cdot q){\hat \mu}_G^2 + \frac{2}{3 m_b^3} \big[ (4m_b-3v \cdot q)\,{\hat \rho}_D^3+ (2m_b-3 v \cdot q){\hat \rho}_{LS}^3 \big]  ,\qquad 
\\
{\tilde S}_2 &=&\frac{7 v \cdot q}{3m_b^2}{\hat \mu}_\pi^2+
\frac{1}{ m_b^2} ( 2m_b-v \cdot q ) {\hat \mu}_G^2 + \frac{2}{3 m_b^3}  ( 4m_b-v \cdot q)\,{\hat \rho}_D^3  ,
\\
{\tilde T}_3 &=&\frac{4}{3 m_b^2}(v \cdot q )^2{\hat \mu}_\pi^2+\frac{4}{3 m_b^3} (m_b-v \cdot q) (v \cdot q ) (2{\hat \rho}_D^3+{\hat \rho}_{LS}^3)  ,
\\
{\tilde S}_3 &=&\frac{4}{3 m_b^2}(v \cdot q )^2{\hat \mu}_\pi^2+\frac{8}{3 m_b^3} (m_b-v \cdot q) (v \cdot q ) \, {\hat \rho}_D^3  ,
\\
{\tilde T}_4 &=&\frac{8}{3 m_b^3} (m_b-v \cdot q) (v \cdot q)^2{\hat \rho}_D^3  , \\
{\tilde S}_4 &=&\frac{8}{3 m_b^3} (m_b-v \cdot q) (v \cdot q)^2{\hat \rho}_D^3  .
\eea
Using these results,  from Eq.~(\ref{dgamma})  the distribution $\displaystyle\frac{d^2 \Gamma}{d y \, d \cos \theta_P}$ in the photon energy $y = 2 \, E_\gamma / m_b$ and  $\cos \theta_P$ can be computed,   $\theta_P$ being  the angle between the photon momentum $\vec q$  and the $H_b$  spin vector $\vec s$ in the $H_b$ rest frame. Upon integration,   the photon energy spectrum, the  $\cos \theta_P$ distribution  and  the decay width are obtained.

In the $H_b$ rest frame  $v \cdot q = E_\gamma = \displaystyle\frac{m_b}{2} \, y $,    and the distribution comprises two terms: 
\bea
\frac{d^2 \Gamma}{dy \, d \cos \theta_P}&=& {\tilde \Gamma}_1+\cos \theta_P \,\,{\tilde \Gamma}_2   \,\,.\label{double}
\eea
Integrating \eqref{double} on $\cos \theta_P$ one has
${\tilde \Gamma}_1=\displaystyle \frac{1}{2} \frac{d \Gamma}{dy }$ and
the photon energy spectrum
\bea
\frac{1}{\Gamma_0} \, \frac{d \Gamma}{d y} & = &\bigg[ 1 - \frac{\hat{\mu}_\pi^2}{2 \, m_b^2} - \frac{\hat{\mu}_G^2}{2 \, m_b^2} \, \frac{3 + 5 \, z}{1 - z} - \frac{10 \, \hat{\rho}_D^3}{3 \, m_b^3} \, \frac{1 + z}{1 - z} \bigg] \, \delta ( 1 - z - y )  \nn \\
& + & \bigg[ \frac{\hat{\mu}_\pi^2}{2 \, m_b^2} \, ( 1 - z ) - \frac{\hat{\mu}_G^2}{6 \, m_b^2} \, ( 3 + 5 \, z ) - \frac{4 \, \hat{\rho}_D^3}{3 \, m_b^3} \, ( 1 + 2 \, z ) + \frac{2 \, \hat{\rho}_{LS}^3}{3 \, m_b^3} \, ( 1 + z ) \bigg] \, \delta' ( 1 - z - y )  \quad \nn \\
& + & \bigg[ \frac{\hat{\mu}_\pi^2}{6 \, m_b^2} \, ( 1 - z )^2 - \frac{\hat{\rho}_D^3}{3 \, m_b^3} \, ( 1 - z ) \, ( 1 + 2 \, z ) + \frac{\hat{\rho}_{LS}^3}{6 \, m_b^3} \, ( 1 - z^2 ) \bigg] \, \delta'' ( 1 - z - y )  \label{photonspectrum} \\
&  - & \frac{\hat{\rho}_D^3}{18 \, m_b^3} \, ( 1 - z )^2 \, ( 1 + z ) \, \delta''' ( 1 - z - y ) \;, \nn 
\eea
%
where $z=\displaystyle\frac{m_s^2}{m_b^2}$ and
\be\displaystyle \Gamma_0 = \frac{\alpha \, G_F^2 \, | \lambda_t |^2}{32 \, \pi^4} \, m_b^5 \, ( 1 - z )^3 \, \left[ | C_+^{\rm eff} |^2 + | C_+^{\prime \, \rm eff} |^2 \right]\,,
\ee
with
\begin{equation*}
C_+^{\rm eff} = C_7^{\rm eff} + \sqrt{z} \, C_7^{\prime \,\rm eff}\,\,, \qquad  \qquad C_+^{\prime \, \rm eff} = \sqrt{z} \, C_7^{\rm eff} + C_7^{\prime \,\rm eff} \;.
\end{equation*}
 ${\tilde \Gamma}_2  $ is given by:
\begin{align}
\frac{2}{\Gamma_0} \tilde{\Gamma}_2 & = -   \frac{| C_+^{\rm eff} |^2 - | C_+^{\prime \, \rm eff} |^2}{| C_+^{\rm eff} |^2 + | C_+^{\prime \, \rm eff} |^2} \,  \notag \\
& \qquad \times \bigg\{ \bigg[ 1 - \frac{13 \, \hat{\mu}_\pi^2}{12 \, m_b^2} - \frac{3 \, \hat{\mu}_G^2}{4 \, m_b^2} \, \frac{5 + 3 \, z}{1 - z} - \frac{\hat{\rho}_D^3}{6 \, m_b^3} \, \frac{31 + 9 \, z}{1 - z} \bigg] \, \delta ( 1 - z - y )  \notag \\
& \qquad + \bigg[ \frac{\hat{\mu}_\pi^2}{2 \, m_b^2} \, ( 1 - z ) - \frac{\hat{\mu}_G^2}{2 \, m_b^2} \, ( 3 + z ) - \frac{2 \, \hat{\rho}_D^3}{m_b^3} \, ( 1 + z ) \bigg] \, \delta' ( 1 - z - y )   \\
& \qquad + \bigg[ \frac{\hat{\mu}_\pi^2}{6 \, m_b^2} \, ( 1 - z )^2 - \frac{\hat{\rho}_D^3}{3 \, m_b^3} \, ( 1 - z ) \, ( 1 + 2 \, z ) \bigg] \, \delta'' ( 1 - z - y )   \notag \\
& \qquad - \frac{\hat{\rho}_D^3}{18 \, m_b^3} \, ( 1 - z )^2 \, ( 1 + z ) \, \delta''' ( 1 - z - y ) \bigg\} \;.\notag
\end{align}
The   $\cos \theta_P$   distribution  also comprises two terms:
\begin{equation}
\frac{d \Gamma (H_b \to X_s \, \gamma)}{d \cos \theta_P}  = A + B \, \cos \theta_P \,\, , \label{angular}
\end{equation}
with
\begin{align}
A & = \frac{1}{2} \, \Gamma (H_b \to X_s \, \gamma) \;, \label{A}\\
B & = -  \frac{\Gamma_0}{2} \, \frac{| C_+^{\rm eff} |^2 - | C_+^{\prime \, \rm eff} |^2}{| C_+^{\rm eff} |^2 + | C_+^{\prime \, \rm eff} |^2} \,  \bigg[ 1 - \frac{13 \, \hat{\mu}_\pi^2}{12 \, m_b^2} - \frac{3 \, \hat{\mu}_G^2}{4 \, m_b^2} \, \frac{5 + 3 \, z}{1 - z} - \frac{\hat{\rho}_D^3}{6 \, m_b^3} \, \frac{31 + 9 \, z}{1 - z} \bigg] \;.\label{B}
\end{align}
%
The  inclusive  $H_b \to X_s \, \gamma$ decay width  is given by
\begin{equation}
\Gamma (H_b \to X_s \, \gamma)  = \Gamma_0 \, \bigg[ 1 - \frac{\hat{\mu}_\pi^2}{2 \, m_b^2} - \frac{\hat{\mu}_G^2}{2 \, m_b^2} \, \frac{3 + 5 \, z}{1 - z} - \frac{10 \, \hat{\rho}_D^3}{3 \, m_b^3} \, \frac{1 + z}{1 - z} \bigg]\,\, . \label{totalG}
\end{equation}
 The SM  result is recovered for $C_7^\prime \to 0$.  At ${\cal O}(1/m_b^2)$ and for $m_s=0$ Eqs.~(\ref{totalG}) and (\ref{photonspectrum})  agree  with the SM expressions  obtained in  \cite{Falk:1993dh, Kapustin:1995nr}. 
 At ${\cal O}(1/m_b^3)$ they agree  with the expressions in \cite{Bauer:1997fe} substituting
\bea
&&
\hat{\mu}_\pi^2 \to - \Big( \lambda_1 + \frac{{\cal T}_1 + 3 \, {\cal T}_3}{m_b} \Big) \; \nn \\
&&
\hat{\mu}_G^2 \to 3 \, \Big( \lambda_2 + \frac{{\cal T}_3 + 3 \,  {\cal T}_4}{m_b} \Big) - \frac{\rho_1 + 3 \, \rho_2}{m_b} \;  \label{subs}\\
&&
\hat{\rho}_D^3 \to \rho_1 \;  \nn \\
&&
\hat{\rho}_{LS}^3 \to 3 \, \rho_2 \;. \nn
\eea
For $m_s=0$ the  ${\cal O}(1/m_b^4)$ corrections  to the decay width have been computed   in \cite{Mannel:2018mqv}.
For $\Lambda_b$,   the distribution ~(\ref{angular}) has been computed at ${\cal O}(1/m_b^2)$  \cite{Gremm:1995nx},
 we agree with such a result. 
 
\section{Photon polarization}\label{sec:polarization}
The photon polarization in the $b \to s \gamma$ transition can be measured in radiative beauty baryon decays. 
In SM the photon polarization asymmetry $A_P$, measuring the relative abundance of left-handed with respect to right-handed photons, is predicted  $A_P \simeq -1$.  Deviations  from this result would  hint  physics  beyond the Standard Model.
The photon polarization  has been studied for the  exclusive \cite{Mannel:1997xy,Huang:1998ek,Hiller:2001zj,Colangelo:2007jy,Oliver:2010im,Blake:2015tda} and inclusive  beauty baryon decays \cite{Gremm:1995nx},  the experimental feasibility  at LHCb has also been scrutinized \cite{GarciaMartin:2019bxm}. 

The correlation between the photon polarization and the initial $b$-baryon polarization is particularly relevant, and it can be obtained  considering the $A_P$ dependence  on $\cos \theta_P$.  We  refer  to $\Lambda_b$, however  our results can be applied also to other beauty baryons.  Preliminary experimental analyses  for $\Sigma_b$ are reported in  \cite{LHCb:2021hfz}.

To present the results for the   photon polarizations $\epsilon_\pm$
we write the double differential width as
\bea
\frac{d^2 \Gamma_\pm}{d y \, d \cos \theta_P} &=& {\tilde \Gamma}_{\pm ,\,1} + {\tilde \Gamma}_{\pm ,\,2} \, \cos \theta_P \; , \label{double_pm}
\eea
where
${\tilde \Gamma}_{\pm , \,1}=\displaystyle \frac{1}{2} \frac{d \Gamma_\pm}{dy }$.
The energy spectrum for definite photon helicities has the expression
\bea
\frac{d \Gamma_+}{d y} &=& R_{+,\,1} \, \frac{d \Gamma}{d y} = \frac{| C_+^{\prime \, \rm eff} |^2}{| C_+^{\rm eff} |^2 + | C_+^{\prime \, \rm eff} |^2} \, \frac{d \Gamma}{d y} \label{spettro+} \;, \\
\frac{d \Gamma_-}{d y} &=& R_{-,\,1} \, \frac{d \Gamma}{d y}=\frac{| C_+^{\rm eff} |^2}{| C_+^{\rm eff} |^2 + | C_+^{\prime \, \rm eff} |^2} \, \frac{d \Gamma}{dy} \label{spettro-} \;.
\eea
${\tilde \Gamma}_{\pm ,\,2}$ are given by
\bea
{\tilde \Gamma}_{+ ,\,2} &=& R_{+,\,2} \, {\tilde \Gamma}_2 = - \frac{| C_+^{\prime \, \rm eff} |^2}{| C_+^{\rm eff} |^2 - | C_+^{\prime \, \rm eff} |^2}  \, {\tilde \Gamma}_2 \;,  \label{dG2+}\\
{\tilde \Gamma}_{- ,\,2} &=& R_{-,\,2} \, {\tilde \Gamma}_2 =  \frac{| C_+^{\rm eff} |^2}{| C_+^{\rm eff} |^2 - | C_+^{\prime \, \rm eff} |^2} \, {\tilde \Gamma}_2 \; . \label{dG2-}
\eea
In SM (for $C_7^{\prime \,\rm eff} \to 0$) we have:
\bea
\frac{d \Gamma_+^\text{SM}}{d y} &=& \frac{z}{1 + z} \,\frac{d \Gamma^\text{SM}}{d y} \, , \\
\frac{d \Gamma_-^\text{SM}}{dy} &=& \frac{1}{1 + z} \, \frac{d \Gamma^\text{SM}}{d y} \, ,
\eea
and 
\bea
{\tilde \Gamma}_{+ ,\,2}^\text{SM} &=& - \frac{z}{1 - z} \,{\tilde \Gamma}_2^\text{SM} \, , \\
{\tilde \Gamma}_{- ,\,2} ^\text{SM} &=& \frac{1}{1 - z} \, {\tilde \Gamma}_2^\text{SM} .
\eea
In  SM for $m_s=0$ only the polarization $\epsilon_-$  contributes.
Indeed, the operator $O_7$ produces a right-handed $b$ quark; the massless $s$ quark  has fixed helicity and, due to the angular momentum conservation, the $b$ and $s$ quarks  have spins aligned to the photon spin, in the  opposite direction.

\begin{figure}[!t]
\begin{center}
\includegraphics[width = 0.6\textwidth]{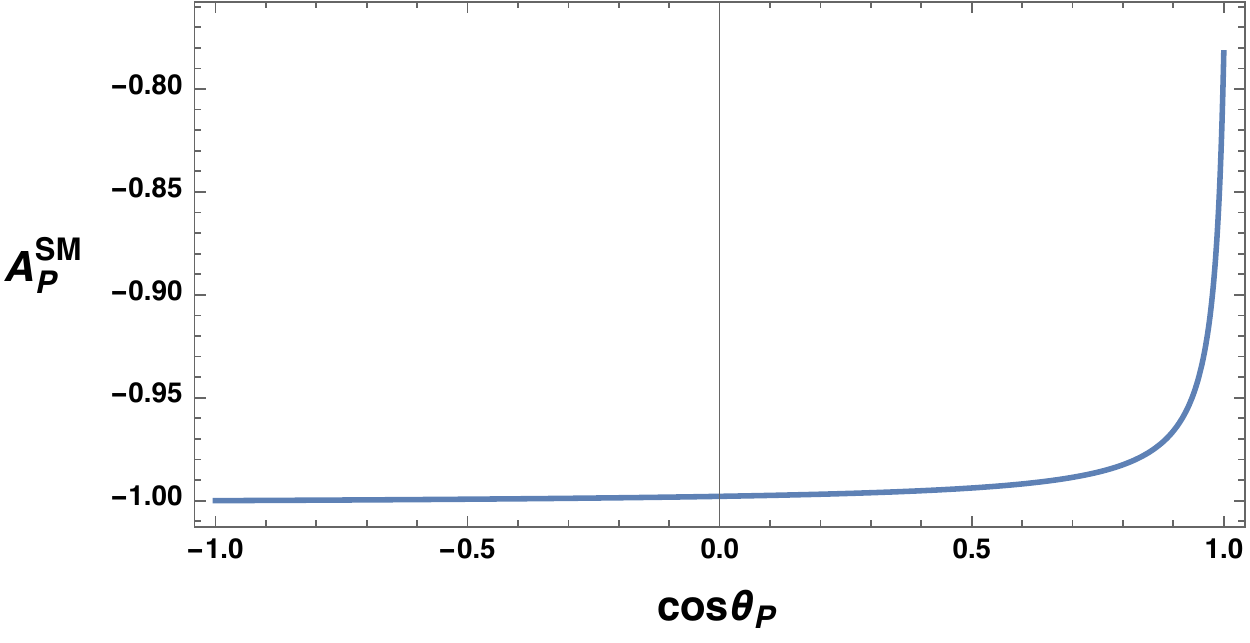}
\caption{\small  Photon polarization asymmetry  Eq.~(\ref{AP}) versus $\cos \theta_P$ in SM.}\label{APSM}
\end{center}
\end{figure}

\begin{figure}[!t]
\begin{center}
\includegraphics[width = 0.55\textwidth]{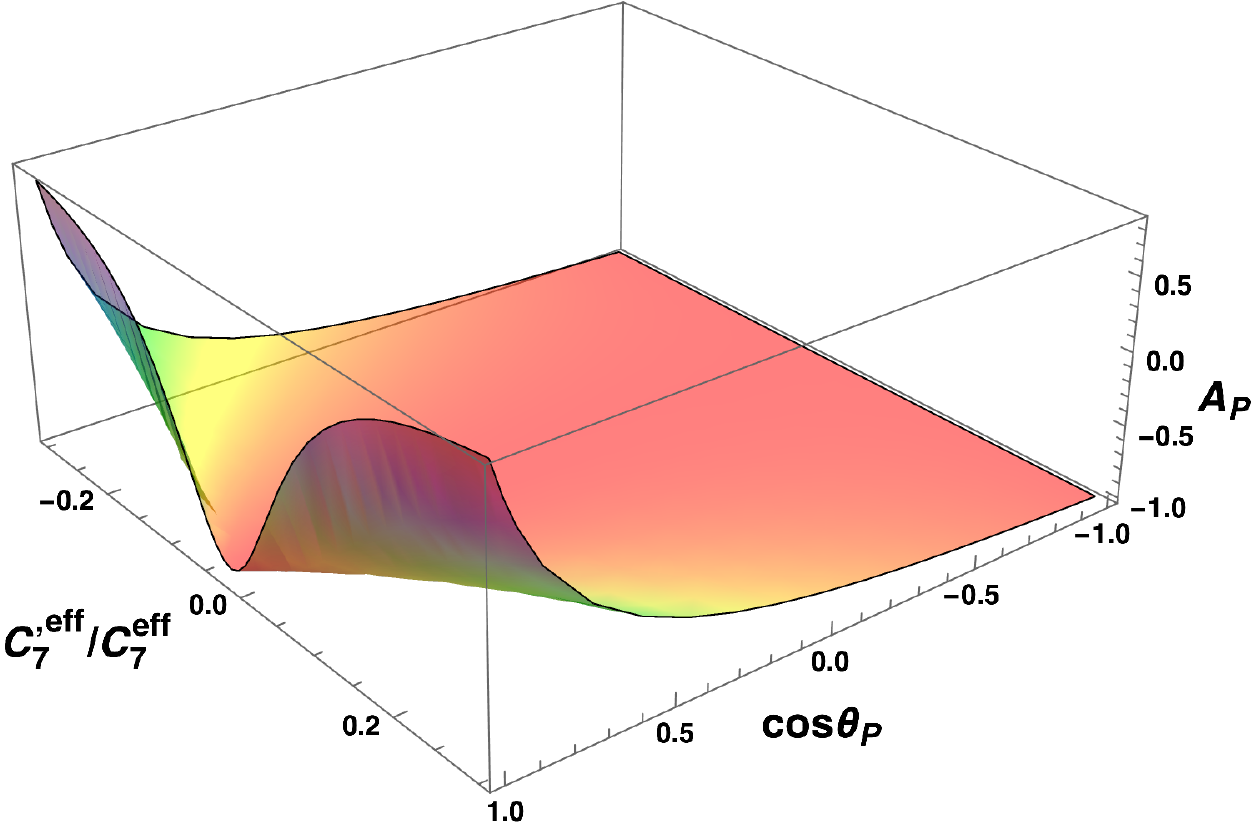}\\ \vspace{0.4cm} \hspace{0.5cm}
\includegraphics[width = 0.7\textwidth]{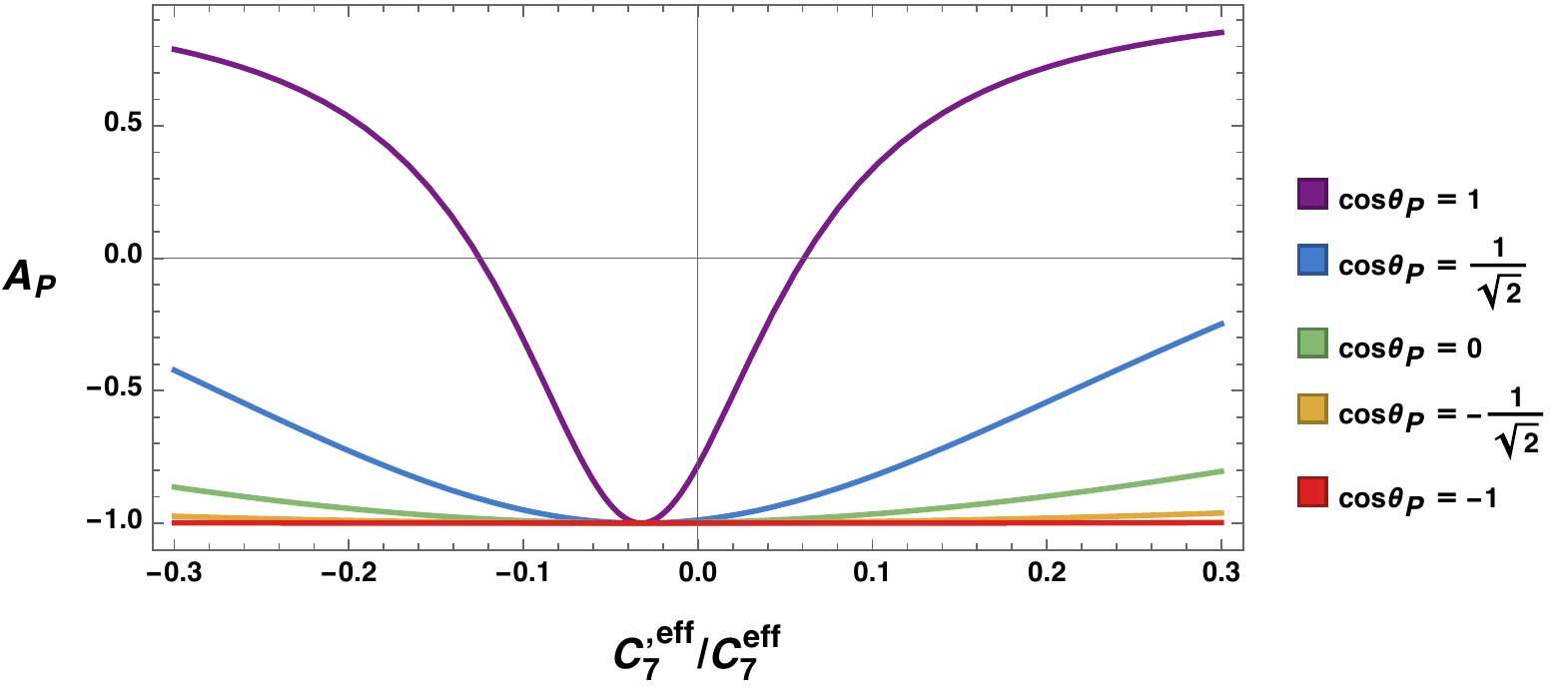}
\caption{\small  Photon polarization asymmetry  Eq.~(\ref{AP}) varying $\cos \theta_P$ and $C^{\prime \, {\rm eff}}_7 / C_7^{\rm eff}$ (top panel) and projected for several values of $\cos \theta_P $ (bottom panel).}\label{AP3D}
\end{center}
\end{figure}

For polarized photons  the  $\cos \theta_P$ distributions have the form
\begin{equation}
\frac{d \Gamma_\pm (H_b \to X_s \, \gamma)}{d \cos \theta_P} = A_\pm + B_\pm \, \cos \theta_P \;, \label{angularpm}
\end{equation}
where 
\bea
A_\pm & = &\frac{\Gamma_\pm (H_b \to X_s \, \gamma)}{2} = R_{\pm,\,1} \,A \; , \nn \\
B_\pm & = &R_{\pm , \,2} \, B \; ,
\eea
and $R_{\pm,1(2)}$  defined  in Eqs.~(\ref{spettro+})-(\ref{dG2-}).
The decay widths to polarized photons are given by
\be
\Gamma_\pm (H_b \to X_s \, \gamma)=R_{\pm,\,1} \,\Gamma (H_b \to X_s \, \gamma) . \label{gpol}
\ee
The photon polarization asymmetry is defined as
\be
A_P(\cos \theta_P) = \frac{\displaystyle{\frac{d \Gamma_+}{d \cos \theta_P}} - \displaystyle{\frac{d \Gamma_-}{d \cos \theta_P}}}{\displaystyle{\frac{d \Gamma_+}{d \cos \theta_P}} + \displaystyle{\frac{d \Gamma_-}{d \cos \theta_P}}} \;. \label{AP}
\ee
In SM the photon polarization asymmetry is $A_P(\cos \theta_P) \simeq - 1$ for almost  all  $\cos \theta_P$,  it  increases only for $\cos \theta_P \to 1$,  see Fig.~\ref{APSM}.
 Physics beyond SM can produce a sizable effect. For a  quantitative insight on the possible deviation from SM,  we consider   ranges for $C_7^\text{NP}=C_7^{\rm eff} -(C_7^{\rm eff})^\text{SM}$ and  $C_7^{\prime \,\rm eff}$,  assuming  that both coefficients are real,  exploiting  the results of a global fit of the  $b \to s $ transitions  \cite{Paul:2016urs}. 
Using $m_b = 4.62 \, \text{GeV}$, $m_s = 0.150 \, \text{GeV}$, ${\hat \mu}_\pi^2(\Lambda_b)= 0.5 \, \text{GeV}^2$, ${\hat \rho}_D^3(\Lambda_b)= 0.17 \, \text{GeV}^3$, ${\hat \mu}_G^2(\Lambda_b) = {\hat \rho}_{LS}^3(\Lambda_b) = 0$, and
varying   $C^{\prime \,\rm eff}_7 / C_7^{\rm eff}  \in [-0.3,0.3]$ we obtain the asymmetry shown in Fig.~\ref{AP3D}.   In the same figure  we plot  $A_P(\cos \theta_P)$ versus  $C^{\prime \,\rm eff}_7 / C_7^{\rm eff} $ for selected values of $\cos \theta_P$.
A  deviation of  the polarization asymmetry from the SM value can be obtained,  with the largest effect   for $\cos \theta_P \simeq 1$.


\section{Treatment of the  singular terms}\label{sf}
 The spectrum obtained by the short distance OPE  does not account for the Fermi motion of the  $b$ quark  due to  soft interactions with the light degrees of freedom in the hadron. For decays of a beauty hadron $H_b$  to light partons the relevant scales  are  $m_b$, $\sqrt{m_b \Lambda_{QCD}}$ and $\Lambda_{QCD}$ \cite{Bauer:2003pi}. 
The "shape function" region is the kinematic  region where the hadronic  invariant mass $p_X^2 \sim   {\cal O}(m_{b} \Lambda_{\rm QCD})$. 
Suitable effective field theory must be used in the various regions:  QCD is first matched to the soft collinear effective theory (SCET) \cite{Bauer:2000yr,Bauer:2001ct,Beneke:2002ph,Beneke:2002ni},   followed by matching to HQET \cite{Bauer:2003pi,Bosch:2004th}.  In the shape function region the short distance OPE is replaced by a twist expansion, with  the infinite set of power corrections  resummed into nonperturbative functions.\footnote{The literature on the shape function, on its properties and on the RGE evolution  is wide,  a list of references  is in   \cite{Gambino:2020jvv}.} 

At leading order in the HQE there is a single   shape function  defined in HQET.
Considering the process $H_b \to X_s \gamma$,   one defines the spectral function  $S_s(y)$   \cite{Neubert:1993um}
\be S_s(y)=\langle \delta \left[ 1-y -z+{2 \over m_b} (v-{\hat q}) \cdot i D \right] \rangle , \label{ss}
 \ee
with ${\hat q}=q/m_b^2$. 
 For a generic operator ${\cal O}$   the matrix element in (\ref{ss}) is defined as
\be
 \langle {\cal O} \rangle ={\langle H_b(v)|{\bar h}_v {\cal O}
h_v |H_b(v)\rangle  \over \langle H_b(v)|{\bar h}_v  h_v |H_b(v)\rangle} \,\, , 
 \ee
with $h_v$  the HQET field  with velocity $v$.
Introducing the  vector $n_\mu+\delta n_\mu=2(v-{\hat q})_\mu |_{y=1-z}$, with
 $n^2=0$,  $v \cdot n=1$,  and $ n \cdot \delta n \sim O(\Lambda_{QCD}/m_b)$ in the shape function region \cite{Mannel:1994pm}, and defining $k_+=n \cdot k$  for a vector $k$, we have
\be S_s(y)=\langle \delta \left[ 1-y -z+{i D_+ \over m_b}  \right]
\rangle \,\,.\label{ss1}
 \ee
$S_s(y)$ can be expressed as 
\be S_s(y)=\int dk_+ \delta \left(1-y-z+{k_+ \over m_b}\right)[f(k_+)+ {\cal O}(m_b^{-1})] \, . \label{intfk} 
\ee 
The function
\be f(k_+)= \langle \delta(i D_+-k_+) \rangle \label{fkpiu} \,\,\ee
 is the  leading shape function.\footnote{Perturbative corrections to the shape function and  to its moments are discussed in \cite{Bauer:2003pi,Bosch:2004th}.}
 The photon energy spectrum  is given by the convolution \cite{Neubert:1993um}
\be
\frac{d \Gamma}{dy}=\int dk_+ f(k_+) \frac{d \Gamma}{dy}^* \, .  \label{convolution}
\ee
In $\displaystyle\frac{d \Gamma}{dy}^*$    the $b$ quark mass $m_b$  is replaced by $m_b^*=m_b+k_+$, an exact  substitution at tree level. For $k_+$  in the range $k_+ \in [-m_b,\,m_{H_b}-m_b]$,   replacing $m_b \to m_b^*$ in  the variable $y$,  we find (for $m_s=0$ to simplify the discussion)  that  $y \to \displaystyle\frac{ 2E_\gamma}{(m_b+k_+)}$.  
Therefore,  for $k_+^{\rm max}=m_{H_b}-m_b$ the maximum photon energy is $E_\gamma=\displaystyle\frac{m_{H_b}}{2}$,  the physical endpoint. 

The  shape function provides an  interpretation of the singular terms in the photon energy spectrum obtained  in the previous Sections. 
The   distribution  in \eqref{photonspectrum} can be written as
\be
\frac{1}{\Gamma} \, \frac{d \Gamma}{d y}=\sum_{n = 0}^3\, \frac{M_n}{n!} \, \delta^{(n)} ( 1 - z - y ) , \label{dGdy}
\end{equation}
with
$\Gamma$  in Eq.~\eqref{totalG} and   $M_{0,\dots 3}$ given by
\bea
M_0 &=&1  \,, \nn
\\
M_1 &=& \frac{\hat{\mu}_\pi^2}{2 \, m_b^2} \, ( 1 - z ) - \frac{\hat{\mu}_G^2}{6 \, m_b^2} \, ( 3 + 5 \, z ) - \frac{4 \, \hat{\rho}_D^3}{3 \, m_b^3} \, ( 1 + 2 \, z ) + \frac{2 \, \hat{\rho}_{LS}^3}{3 \, m_b^3} \, ( 1 + z ) \,, \nn \\
M_2 &=&  \frac{\hat{\mu}_\pi^2}{3 \, m_b^2} \, ( 1 - z )^2 - \frac{2\hat{\rho}_D^3}{3 \, m_b^3} \, ( 1 - z ) \, ( 1 + 2 \, z ) + \frac{\hat{\rho}_{LS}^3}{3 \, m_b^3} \, ( 1 - z^2 ) \label{momspectrumSF}\, ,  \\
M_3 &=& - \frac{\hat{\rho}_D^3}{3 \, m_b^3} \, ( 1 - z )^2 \, ( 1 + z )\,. \nn
\eea
At ${\cal O}(1/m_b^2)$  Eqs.~\eqref{momspectrumSF} agree with the  expressions obtained  in \cite{Neubert:1993um}.
They can be considered as  the first few terms of the infinite sum \cite{Bigi:1993fe,Bigi:1993ex,Neubert:1993ch,Neubert:1993um,Mannel:1994pm}
\be
S_s(y)=\sum_{n = 0}^{\infty} \, \frac{M_n}{n!} \, \delta^{(n)} ( 1 - z - y ) \; . \label{SF}
\end{equation}
As pointed out in \cite{Neubert:1993um},  a feature of Eqs.~\eqref{momspectrumSF} is that  each moment $M_n$ has an expansion in powers of  $1/m_b$   starting at the same order of the  moment,
 \be
M_n = \sum_{k=n}^\infty \displaystyle\frac{M_{n,k}}{m_b^k}\,.
\ee
Analogously, the double differential distribution \eqref{double} can be written as
\be
\frac{d^2 \Gamma}{d y \, d \cos \theta_P} = A \,\Bigg[ \sum_{n = 0}^3\, \frac{M_n}{n!} \, \delta^{(n)} ( 1 - z - y ) \Bigg]
+B \, \,\Bigg[\sum_{n = 0}^3\, \frac{M_n^{\theta_P}}{n!} \, \delta^{(n)} ( 1 - z - y ) \Bigg]\, \cos \theta_P \;, \label{ddd} 
\ee
with $A,\,B$   in \eqref{A}, \eqref{B}. 
The moments $M_n$   in Eq.~\eqref{photonspectrum}  can be considered as resulting from the expansion of the function \eqref{SF}, and  the terms
\bea
M_0^{ \theta_P} &=& 1  \nn \\
M_1^{ \theta_P} &=&\frac{\hat{\mu}_\pi^2}{2 \, m_b^2} \, ( 1 - z ) - \frac{\hat{\mu}_G^2}{2 \, m_b^2} \, ( 3 + z ) - \frac{2 \, \hat{\rho}_D^3}{m_b^3} \, ( 1 + z ) \nn \\
M_2^{ \theta_P} &=&\frac{\hat{\mu}_\pi^2}{3 \, m_b^2} \, ( 1 - z )^2 - \frac{2 \, \hat{\rho}_D^3}{3 \, m_b^3} \, ( 1 - z ) \, ( 1 + 2 \, z ) \label{smeno} \\
M_3^{ \theta_P} &=&- \frac{\hat{\rho}_D^3}{3 \, m_b^3} \, ( 1 - z )^2 \, ( 1 + z )   \; \nn
\eea
 as deriving from the function
\begin{equation}
\label{spiumeno}
S_s^{ \theta_P}(y) = \sum_{n = 0}^{\infty} \, \frac{M_n^{ \theta_P}}{n!} \, \delta^{(n)} ( 1 - z - y ) \;.
\end{equation}
The  factors $M_n$ in \eqref{SF} are related to the moments of the photon energy spectrum
\be
\langle y^k \rangle  = \frac{1}{\Gamma} \, \int_0^{1 - z} \, d y \, y^k \, \frac{d \Gamma}{d y}  \, . 
\ee
Indeed, using \eqref{dGdy} we have:
\begin{equation}\label{moments}
\langle y^k \rangle  =   \sum_{n = 0}^\infty \, \frac{M_n}{n!} \, \int_0^{1 - z} \, d y \, y^k \, \delta^{(n)} ( 1 - z - y )  
 =   \sum_{j = 0}^k \, \binom{k}{j} \, ( 1 - z )^{k - j} \,M_j  \, , 
\end{equation}
 and
\bea
\langle y \rangle & =  &( 1 - z ) \, + M_1  \;, \label{stat1}\\
\langle y^2 \rangle & = & ( 1 - z )^2 \,  + 2 \, ( 1 - z ) \, M_1 + M_2  \;, \label{stat2}\\
\sigma_y^2& = & \langle y^2 \rangle - \langle y \rangle^2 =   M_2 - M_1^2 \;. \label{stat3}
\eea
Such results imply that
\bea
\langle y \rangle & = & ( 1 - z ) \, \bigg[ 1 + \frac{\hat{\mu}_\pi^2}{2 \, m_b^2} - \frac{\hat{\mu}_G^2}{6 \, m_b^2} \, \frac{3 + 5 \, z}{1 - z} - \frac{4 \, \hat{\rho}_D^3}{3 \, m_b^3} \, \frac{1 + 2 \, z}{1 - z} + \frac{2 \, \hat{\rho}_{LS}^3}{3 \, m_b^3} \, \frac{1 + z}{1 - z} \bigg] \;, \label{y}\\
\langle y^2 \rangle & = & ( 1 - z )^2 \, \bigg[ 1 + \frac{4 \, \hat{\mu}_\pi^2}{3 \, m_b^2} - \frac{\hat{\mu}_G^2}{3 \, m_b^2} \, \frac{3 + 5 \, z}{1 - z} - \frac{10 \, \hat{\rho}_D^3}{3 \, m_b^3} \, \frac{1 + 2 \, z}{1 - z} + \frac{5 \, \hat{\rho}_{LS}^3}{3 \, m_b^3} \, \frac{1 + z}{1 - z} \bigg] \;, \label{y2}\\
\sigma_y^2 & = & ( 1 - z )^2 \, \bigg[ \frac{\hat{\mu}_\pi^2}{3 \, m_b^2} - \frac{2 \, \hat{\rho}_D^3}{3 \, m_b^3} \, \frac{1 + 2 \, z}{1 - z} + \frac{\hat{\rho}_{LS}^3}{3 \, m_b^3} \, \frac{1 + z}{1 - z} \bigg] \; .\label{sig}
\eea
After the substitutions  in \eqref{subs}, the above expressions agree with those given in \cite{Benson:2004sg} for  $z=0$.

\begin{table}[!t]
\center
\begin{tabular}{|c|c|c|c|}
\cline{2-4}
\multicolumn{1}{c|}{} & LO & $O(1 / m_b^2)$ & $O(1 / m_b^3)$ \\
\hline
$\langle y \rangle$ & $0.999$ & $1.011$ & $1.008$ \\
\hline
$\langle y^2 \rangle$ & $0.998$ & $1.029$ & $1.023$ \\
\hline
$\sigma_y^2$ & $0$ & $0.008$ & $0.007$ \\
\hline
\end{tabular}
\smallskip
\caption{First  moments of the photon energy spectrum at LO, ${ O}(1/m_b^2)$, ${O}(1/m_b^3$).}
\label{energy_moments}
\end{table}
Table~\ref{energy_moments} contains numerical results for the first  moments,  increasing the order in $1/m_b$ and using the  parameters  in Section \ref{sec:polarization}.
The moments of the measured photon energy spectrum can be used to determine  the HQET parameters, as for  $B$ mesons.  Baryons have the advantage that the $\cos \theta_P$ distribution \eqref{angular} can also be exploited.

 The  Fermi motion of the heavy quark in the hadron has the effect of smearing the spectrum. Indeed, 
at the various orders in the $1/m_b$ expansion, the photon energy spectrum obtained by the  local OPE corresponds to a monochromatic  line. At the leading order the line is  placed at $y=2E_\gamma/m_b=1-z$,   the next terms  correspond to  a  displacement of this position. The convolution  with the shape function  $f(k_+)$  in  \eqref{convolution}  produces the smearing. The shape function is a nonperturbative quantity,  it must be determined by  methods as lattice QCD or QCD sum rules starting from the moments, or it must be  modelized  to reproduce the experimental photon spectrum \cite{Lange:2005yw,DeFazio:1999ptt,Andersen:2005mj,Gambino:2007rp,Aglietti:2007ik}.  In the latter case, the uncertainty connected to the functional dependence is usually estimated using different model functions, or varying the model parameters.

 Instead of parametrizing the shape function,  we proceed considering that, if an infinite number of terms is included, the sum  Eq.~(\ref{dGdy})  gives the spectral function $S_s(y)$  in \eqref{SF}. 
In the sum the first term corresponds to a monocromatic line at the zero of the $\delta$-function, with  $\langle y \rangle=(1-z)$ and $\sigma_y^2=0$, the leading order results   in Eqs.~\eqref{stat1}-\eqref{stat3}.
We observe that  the Dirac delta can be represented as
\be
\delta(1-z-y)=\lim_{\sigma_y \to 0}\frac{1}{\sqrt{2\pi} \sigma_y} e^{-\frac{(b-y)^2}{2\sigma_y^2}} \, \label{deltalimit}
\ee
with $b=(1-z)=\langle y \rangle_{LO}$. We can fix  $\sigma_y^2$  at each order  in $1/m_b$,  starting from $1/m_b^{2}$.   For $m_b \to \infty$ the limit $\sigma_y \to 0$
reproduces the partonic result.   

We  represent the spectral function $S_s(y)$   by the substitution
\bea
S_s(y)&=&\sum_{n = 0}^{\infty} \, \frac{M_n}{n!} \, \delta^{(n)} ( 1 - z - y ) 
\to S_{s }(y)=\sum_{n = 0}^{\infty} \, \frac{M_n}{n!} \,(-1)^n \frac{d^n}{dy^n}\frac{1}{\sqrt{2\pi} \sigma_y} e^{-\frac{(b-y)^2}{2\sigma_y^2}}\,\, . \label{Ss0}
\eea
Using the representation of the  Hermite polynomials
\be
H_n(x)=(-1)^n e^{x^2} \frac{d^n}{dx^n}e^{-x^2} \, ,
\ee
the substitution gives
\be
 S_{s }(y)=\frac{1}{\sqrt{2\pi} \,\sigma_y}\sum_{n = 0}^{\infty} \, \frac{M_n}{n!}\left(-\frac{1}{\sqrt{2}\sigma_y}\right)^ne^{-\frac{(b-y)^2}{2\,\sigma_y^2}}H_n\left(\frac{b-y}{\sqrt{2}\,\sigma_y}\right) . \label{Ss}
\ee
Notice that, denoting  by $\langle y^k \rangle_{\cal N}$ the moments computed using this expression for the spectral function $S_s(y)$,
\be
\langle y^k \rangle_{\cal N}=\int_0^{y_{\rm max}}\,dy \,  y^k \,  S_{s }(y) \,\,, \label{momN}
\ee 
 in the limit $\sigma_y \to 0$ one obtains
\be
\lim_{\sigma_y \to 0} \langle y^k \rangle_{\cal N}= \langle y^k \rangle  \label{limit}
\ee
with  $\langle y^k \rangle$   in \eqref{moments}.  This is shown in Appendix \ref{appA}.

The ansatz Eq.~\eqref{Ss} has many advantages with respect to other representations of the shape function or of the spectral function, based on a choice of a functional representation able to reproduce the  photon  spectrum, with parameters set  by the first computed moments $M_n$. Indeed, such representations generally do not guarantee that higher moments are reproduced. Moreover,  in such models  the moments $M_n$  
generally increase with the order $n$ \cite{MannelMainz}.   To cure such features, in \cite{Ligeti:2008ac}  the shape function  is expressed using a complete set of orthonormal functions. In particular, the normalized Legendre polynomials are considered in the range $[-1,+1]$, and a  function mapping the range  $[-1,1]$ into $[0,+\infty)$ is chosen to represent the  shape function $f(\omega)$  in the definition having  support  $\omega \in[0,+\infty)$. 

 Remarkably,  with the  ansatz in Eq.~\eqref{Ss}, by construction  $S_s(y)$  can include all moments $M_n$ once they are computed.  Moreover, less singular terms in the expansion of the moments are not discarded. Each $M_n$ starts at ${\cal O}(1/m_b^n)$ and depends on the matrix elements of the HQET operators of increasing dimension, 
\be
{\cal M}_{\mu_1 \dots \mu_n}=\langle H_b(v,s)|({\bar b}_v)_a(i D_{\mu_1})\dots(i D_{\mu_n})(b_v)_b |H_b(v,s)\rangle \label{matel}
\ee  of  ${\cal O}(\Lambda_{QCD}^n)$, maintaining a hierarchical ordering. 
 
We point out the main differences with  the method  proposed in \cite{Ligeti:2008ac} and adopted in \cite{Bernlochner:2020jlt}. In that approaches it is assumed  that the leading shape function is positive, hence it is expressed as the square of the sum of  orthonormal functions, choosing in particular the normalized Legendre polynomials.
The  positivity assumption is not necessary in  the ansatz \eqref{Ss}. In such expression  the expansion in Hermite polynomials is not arbitrary, since it comes from the replacement of  the  Dirac delta  by the normal distribution with standard deviation $\sigma_y$  \eqref{deltalimit}. This produces a different result  from the single gaussian function in e.g. \cite{Altarelli:1982kh,Mannel:1994pm}. Indeed, the derivatives of the exponential in Eq.~\eqref{Ss0} produce coefficients with a  non trivial dependence on $y$, which are resummed giving the Hermite polynomials in \eqref{Ss}. This  produces the asymmetry of the shape function with respect to the point $y=b$.
 While $b$ is fixed to the LO result for $\langle y \rangle$, $\sigma_y$ can be determined at an arbitrary order in the $1/m_b$ expansion;  Eq.\eqref{sig}  satisfies by construction the condition  $\dd \lim_{m_b \to \infty} \sigma_y =0$, recovering the monochromatic spectrum in the limit. In \eqref{Ss} the Hermite polynomials are not weighted by new unkown coefficients to be fitted, but  by the computed moments $M_n$. 
 
\begin{figure}[!ht]
\begin{center}
\includegraphics[width = 0.6\textwidth]{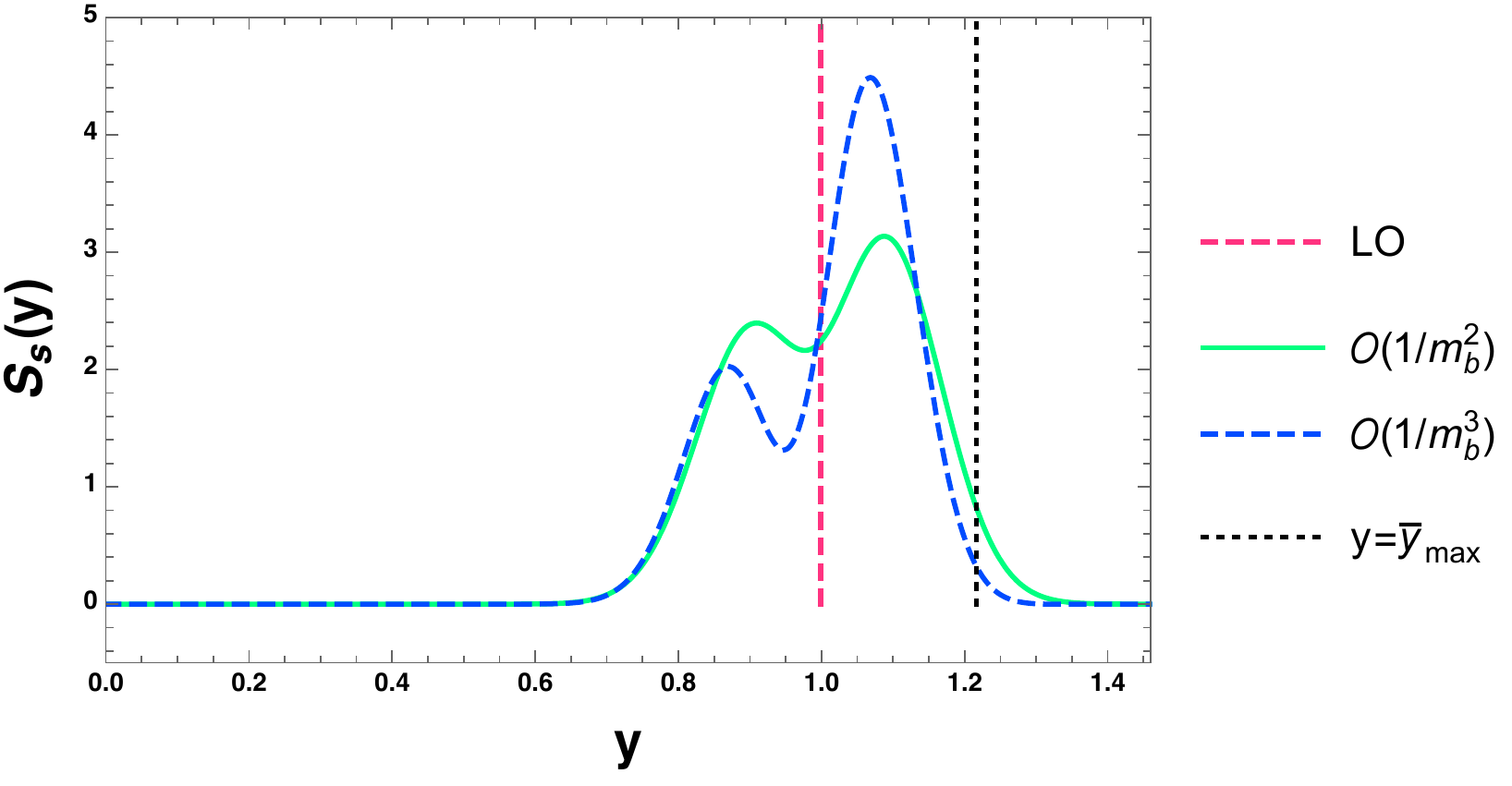}\\
\includegraphics[width = 0.6\textwidth]{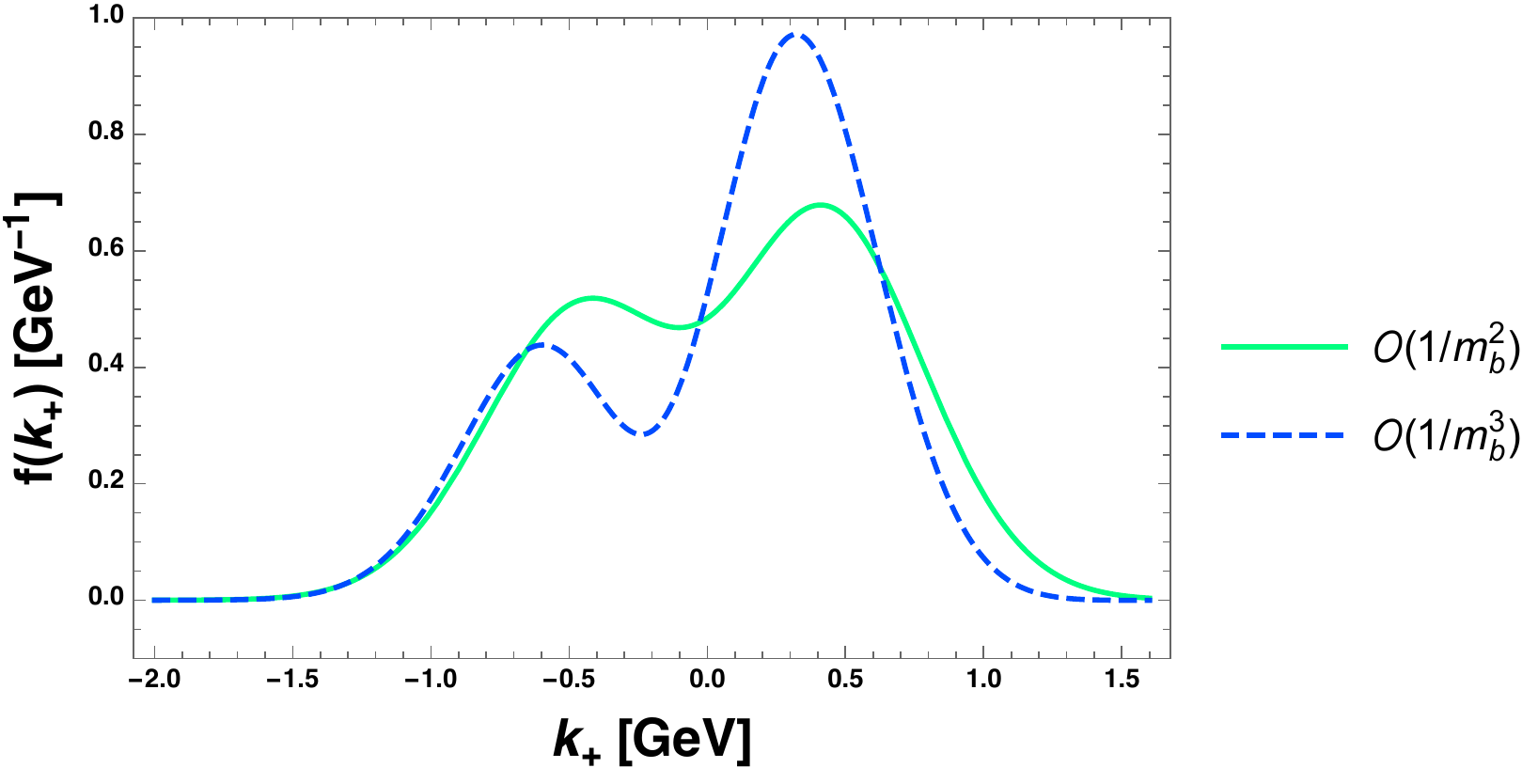}
\caption{\small  Spectral function $S_s(y)$ (top panel) and  shape function $f(k_+)$ (bottom panel) obtained using the ansatz Eq.~\eqref{Ss} up to $n=3$.}\label{spectrumSF}
\end{center}
\end{figure}

For an analysis based on the ansatz \eqref{Ss}, 
in  Fig.~\ref{spectrumSF} we show the spectral function obtained at LO,   ${\cal O}(1/m_b^2)$ and  ${\cal O}(1/m_b^3)$. 
 In the same figure we  plot the shape function obtained from \eqref{intfk}  at     ${\cal O}(1/m_b^2)$ and  ${\cal O}(1/m_b^3)$.
 As a consequence of broadening  the spectrum through the substitution in \eqref{Ss0},  there is a tail  exceeding the physical endpoint  ${\bar y}_{\rm max}\simeq\displaystyle\frac{m_{\Lambda_b}}{m_b}(1-z)$, and  a tail in the shape function exceeding $k_+^{\rm max}=m_{\Lambda_b}-m_b$. This is a spurious effect of the truncation. When higher orders in the HQE are included, the area below such  tails approaches to zero.  Indeed, denoting this  area  by $\dd \Delta \langle y^0 \rangle_n=\int_{{\bar y}_{\rm max}}^\infty dy\,  \frac{1}{\Gamma} \frac{d \Gamma}{dy}$ computed at ${\cal O}(1/m_b^{n})$, we  numerically find  $\displaystyle\frac{\Delta \langle y^0 \rangle_3}{\Delta \langle y^0 \rangle_2}\simeq {\cal O}\left(\frac{1}{m_b} \right)$. Increasing the order in the HQ expansion $\Delta \langle y^0 \rangle_n$ reduces to zero, so that   the physical endpoint is reached.

\section{Conclusions}
The HQE has been exploited to compute the inclusive  decay width induced by the $b \to s \gamma$ transition for a beauty baryon, in particular  $\Lambda_b$.  The  differential width  in the rescaled photon energy $\dd y=\frac{2 E_\gamma}{m_b}$  and in $\cos \theta_P$  allows to construct new observables  with respect to mesons. The calculation has been carried out at ${\cal O}(1/m_b^{3})$ for non-vanishing strange quark mass,  using the baryon matrix elements  determined in  \cite{Colangelo:2020vhu}. Physics beyond the Standard Model   represented by  the operator $O_7^\prime$   is found to affect  the photon polarization asymmetry. 
For  the singular terms appearing in the spectrum as $\delta$-distribution and its derivatives we have proposed a treatment
that can be systematically improved  including higher order terms in the expansion.

Progress in the next studies will be achieved considering the full Hamiltonian \eqref{hamil},  the resolved photon contribution and  the subleading shape functions
for   $b$-baryons.  This is an important step forward, in view of the wealth of new information which can be gained on SM and on  the possible extensions   analyzing the  beauty baryon rare radiative decay modes together with  the   polarization effects.

\section*{Acknowledgements}
We thank Andrzej J. Buras and Gil Paz for discussions.
 This research has  been carried out within the INFN project (Iniziativa Specifica)  QFT-HEP.

\appendix
\section{Derivation of the OPE}\label{newappOPE}
The OPE for the expression \eqref{Tij-gen} can be constructed expressing the hadron momentum $p=m_H v$, with $v$  the four-velocity, in terms of $m_b$ and of a residual momentum $k$: $p=m_b v+k$. The QCD $b$ quark field is rescaled  
\be b(x)=e^{-i\,m_b v \cdot x} b_v(x) ,  \ee
 and the QCD field  $b_v(x)$ satisfies the equation of motion
\be
b_v(x)=\left(P_+ +\frac{i {\spur D}}{2m_b} \right) b_v(x) \,\, ,
\ee
with   velocity projector $P_+=\displaystyle\frac{1+ \spur v}{2}$. Expressed in terms of $b_v(x)$ Eq.~\eqref{Tij-gen} becomes:
\be
T^{ij}_{MN}=i\,\int d^4x \, e^{i\,(m_b v -q) \cdot x} \langle H_b(v,s)|T[ {\hat J}^{i\dagger}_M (x) \, \hat J^{j}_N (0)] |H_b(v,s)\rangle\,\, .
\ee
${\hat J}^{i}$ contains the field $b_v$. 
The heavy quark expansion  is obtained from 
\be
T^{ij}_{MN}=\langle H_b(v,s)|{\bar b}_v(0) {\bar \Gamma}_M^{i} S_s(p_X)\Gamma_N^{j}b_v(0) |H_b(v,s)\rangle\,\, ,
\ee
with $\bar \Gamma^{i}_M=\gamma^0 \Gamma^{i\dagger}_M \gamma^0$ and   $\Gamma_M^{7}=\sigma_{\mu \nu}\,(m_b\,P_R+m_s\,P_L)$, $\Gamma_M^{7^\prime}=\sigma_{\mu \nu}\,(m_b\,P_L+m_s\,P_R)$.  $S_s(p_X)$ is the $s$ quark propagator.  Replacing $k \to iD$, with $D$  the QCD covariant derivative, the $s$ quark propagator  can be expanded:
\be
S_s(p_X)=S_s^{(0)}-S_s^{(0)}(i {\spur D})S_s^{(0)}+S_s^{(0)}(i {\spur D})S_s^{(0)}(i {\spur D})S_s^{(0)}+ \dots\,\, \label{exp-prop}
\ee
where $S_s^{(0)}=\displaystyle\frac{1}{m_b {\spur v}-{\spur q} -m_s}$.
Defining $p_s=m_b v -q$, ${\cal P}={\spur p}_s+m_s$ and $\Delta_0=p_s^2-m_s^2$,  the expansion  at order $1/m_b^3$ is given by:
\bea
\frac{1}{\pi}{\rm Im} \, T^{ij}_{MN}&=& 
\frac{1}{\pi}{\rm Im}\frac{1}{\Delta_0}\langle H_b(v,s)|{\bar b}_v [{\bar \Gamma}_M^{i} {\cal P} \Gamma_N^{j}]b_v |H_b(v,s) \rangle \nn \\
&-&\frac{1}{\pi}{\rm Im}\frac{1}{\Delta_0^2}\langle H_b(v,s)|{\bar b}_v[ {\bar \Gamma}_M^{i} {\cal P}\gamma^{\mu_1}{\cal P} \Gamma_N^{j}](i D_{\mu_1})b_v |H_b(v,s)\rangle \nn \\
&+&\frac{1}{\pi}{\rm Im}\frac{1}{\Delta_0^3}\langle H_b(v,s)|{\bar b}_v[ {\bar \Gamma}_M^{i} {\cal P}\gamma^{\mu_1}
{\cal P}\gamma^{\mu_2}{\cal P} \Gamma_N^{j}](i D_{\mu_1})(i D_{\mu_2})b_v |H_b(v,s)\rangle  \label{expansion}  \\
&-&\frac{1}{\pi}{\rm Im}\frac{1}{\Delta_0^4} \langle H_b(v,s)|{\bar b}_v[ {\bar \Gamma}_M^{i} {\cal P}\gamma^{\mu_1}
{\cal P}\gamma^{\mu_2}{\cal P} \gamma^{\mu_3}{\cal P}\Gamma_N^{j}](i D_{\mu_1})(i D_{\mu_2})(i D_{\mu_3})b_v |H_b(v,s) \rangle \,\,.\nn 
\eea
This expression involves the $H_b$ matrix elements of QCD operators of increasing dimension,  
\bea
&&
\langle H_b(v,s)|{\bar b}_v[ {\bar \Gamma}_M^{i} {\cal P}\gamma^{\mu_1}\dots  \gamma^{\mu_n}{\cal P}\Gamma_N^{j}](i D_{\mu_1})\dots(i D_{\mu_n})b_v |H_b(v,s)\rangle =
\nn  \\
&&{\rm Tr} \left[({\bar \Gamma}_M^{i}{\cal P}\gamma^{\mu_1}\dots  \gamma^{\mu_n}{\cal P}\Gamma_N^{j})_{ba} \langle H_b(v,s)|({\bar b}_v)_a(i D_{\mu_1})\dots(i D_{\mu_n})(b_v)_b |H_b(v,s)\rangle \right] \,  ,  \quad \label{trace-form}
\eea
with $a,b$  Dirac indices. 
The hadronic matrix elements 
\be
\left({\cal M}_{\mu_1 \dots \mu_n}\right)_{ab}=\langle H_b(v,s)|({\bar b}_v)_a(i D_{\mu_1})\dots(i D_{\mu_n})(b_v)_b |H_b(v,s)\rangle \label{matelements}
\ee  
can be expressed in terms of  a set of  nonperturbative parameters, the number of which increases with the operator dimension.
At  ${\cal O}(1/m_b^3)$    the following matrix elements are required:
\bea
\langle H_b(v,s)|{\bar b}_v (iD)^2 b_v|H_b(v,s)\rangle&=&-2m_H\,{\hat \mu}_\pi^2 \label{mupi}\\
\langle H_b(v,s)|{\bar b}_v (iD_\mu)(iD_\nu)(-i \sigma^{\mu \nu})b_v|H_b(v,s)\rangle&=&2m_H\,{\hat \mu}_G^2 \label{mug} 
\\
\langle H_b(v,s)|{\bar b}_v (iD_\mu)(i v \cdot D) (i D^\mu) b_v|H_b(v,s)\rangle&=&2m_H\,{\hat \rho}_D^3\label{rd} \\
\langle H_b(v,s)|{\bar b}_v (iD_\mu)(i v \cdot D) (i D_\nu) (-i \sigma^{\mu \nu})b_v|H_b(v,s)\rangle &=&2m_H\,{\hat \rho}_{LS}^3 \,\, .\label{rls}
 \eea
A procedure to compute  ${\cal M}_{\mu_1 \dots \mu_n}$  has been  exploited for $B$ meson for $n=4$  \cite{Dassinger:2006md} and  $n=5$ \cite{Mannel:2010wj}, introducing additional   parameters  with respect to  those  in (\ref{mupi})-(\ref{rls}). 
For a heavy baryon,
 the dependence on the spin four-vector $s_\mu$   is specified in \eqref{matelements}.
The matrix elements in the expansion at ${\cal O}(1/m_b^3)$ keeping the  $s_\mu$ dependence have been parametrized in \cite{Colangelo:2020vhu} and are used in the present study.

\section{ Factorization formula in the endpoint region and shape functions}\label{newappA}
The local OPE  exploited in this paper holds  in the kinematic region where the hadronic invariant mass $p_X^2$ is ${\cal O}(m_{b}^2)$. In the region where 
$p_X^2 \sim   {\cal O}(m_{b} \Lambda_{\rm QCD})$, the endpoint or  shape function region, the theoretical treatment is different according to which of the operators in the effective Hamiltonian one considers. In our analysis 
we have focused  on the dipole operators $O_7^{(\prime)}$  since they are the only terms contributing to  lowest order in QCD. 
In the SM  
the contribution of $O_7$ to the correlation function \eqref{Tij-gen} at the endpoint obeys a factorization formula   
\be
d\Gamma^{77}\sim H \cdot J \otimes S+\frac{1}{m_b}\sum_i H \cdot J \otimes s_i+\frac{1}{m_b}\sum_i H \cdot J_i \otimes S +{\cal O}\left( \frac{\Lambda_{\rm QCD}^2}{m_b^2}\right)\,\,.
\label{fact}
\ee
$H$ denote hard functions and  $J$ and $j_i$ are jet functions  computed perturbatively, with $H$ of ${\cal O}(1)$.
 $S$ and $s_i$ are the  shape functions, which are nonperturbative. The function $S$ entering in \eqref{fact} at leading order  is the  shape function  defined in HQET. 

When the other operators in \eqref{Tij-gen} are considered, a more involved factorization formula holds at ${\cal O}(1/m_b)$ \cite{Benzke:2010js}.
 The most important  operators are $O_2$ and $O_8$, and the pairing of the  operators in \eqref{Tij-gen} produces different effects. In particular, the resolved photon contribution (RPC) mentioned in the Introduction appears as single RPC from the pairing of  $O_8$ and $O_2$ among themselves and with $O_7$. Double RPC arise from the pairing of $O_8$ and $O_2$ among themselves \cite{Benzke:2010js}.  For such contributions
the leading term is ${\cal O}(\alpha_s)$: this justifies their neglect in the present analysis.

\section{Moments of the photon energy spectrum}\label{appA}
To  obtain  Eq.~\eqref{limit} we use  the representation of the Hermite polynomials
\be
H_n(x)=n! \sum_{m=0}^{\lfloor \frac{n}{2} \rfloor } \frac{(-1)^m}{m!(n-2m)!}(2x)^{n-2m} \, .
\ee
The moments $\langle y^k \rangle_\mathcal{N}$ are given by:
\begin{equation}
\langle y^k \rangle_\mathcal{N}  = \sum_{j = 0}^k \, \binom{k}{j} \, b^{k - j} \, \sum_{n = 0}^\infty \, M_n \, ( - \sqrt{2} \, \sigma_y )^{j - n} \, \Phi_{j,n} \,\, , \label{app:2}
\end{equation}
where
\begin{equation}
\begin{split}
\Phi_{j,n}=&
\frac{1}{\sqrt{\pi}} \, \sum_{m = 0}^{\lfloor \frac{n}{2} \rfloor} \, \frac{( - 1 )^m}{m! \, ( n - 2 \, m )!} \, 2^{n - 2 \, m}   \label{phijn}\\
& \times \frac{1}{2} \, \bigg[ \gamma \bigg( \frac{j + n - 2 \, m + 1}{2}, x_{\max}^2 \bigg) - 
 ( - 1 )^{j + n - 2 \, m + 1} \,  \gamma \bigg( \frac{j + n - 2 \, m + 1}{2}, x_{\min}^2 \bigg) \bigg] \,\, .
\end{split}
\end{equation}
The parameters $b$ and $\sigma_y$ are defined in Section \ref{sf}, and  $x_{\max(\min)}$ are  $x_{\max} = \displaystyle\frac{b }{\sqrt{2} \, \sigma_y}>0$, 
$x_{\min} = \displaystyle\frac{b-y_{\max} }{\sqrt{2} \, \sigma_y}<0$. 
$\gamma(n,z)$  is the lower incomplete Euler function with the condition $\text{Re}(j + n - 2 \, m) > - 1$  always satisfied in our case.
$\Phi_{j,n}$ depends on $\sigma_y$  only through $x_{\max(\min)}$. Since $\displaystyle{\lim_{\sigma_y \to 0}{x_{\max(\min)}} =+(-) \infty}$, we have  $\displaystyle{\lim_{\sigma_y \to 0}{\Phi_{j,n}}=0}$ for $n\neq j$. Consequently, for ${\sigma_y \to 0}$ we obtain that $\langle y^k \rangle_\mathcal{N}$ is  given by the $n = j$ terms in \eqref{app:2},
\be
\langle y^k \rangle_\mathcal{N}  =  \sum_{j = 0}^k \, \binom{k}{j} \, b^{k - j} \, M_j \, \Phi_{j,j} \, .
\end{equation}
Since $\displaystyle{\lim_{\sigma_y \to 0}\Phi_{j,j}=1}$, Eq.~\eqref{moments} is recovered.

\bibliographystyle{JHEP}
\bibliography{refsFFP6}

\providecommand{\href}[2]{#2}\begingroup\raggedright\begin{thebibliography}{10}

\bibitem{Shifman:1976de}
M.~A. Shifman, A.~I. Vainshtein, and V.~I. Zakharov, {\it {On the Weak
  Radiative Decays (Effects of Strong Interactions at Short Distances)}},  {\em
  Phys. Rev. D} {\bf 18} (1978) 2583--2599. [Erratum: Phys.Rev.D 19, 2815
  (1979)].

\bibitem{Bertolini:1986th}
S.~Bertolini, F.~Borzumati, and A.~Masiero, {\it {QCD Enhancement of Radiative
  b Decays}},  {\em Phys. Rev. Lett.} {\bf 59} (1987) 180.

\bibitem{Grinstein:1987vj}
B.~Grinstein, R.~P. Springer, and M.~B. Wise, {\it {Effective Hamiltonian for
  Weak Radiative B Meson Decay}},  {\em Phys. Lett. B} {\bf 202} (1988)
  138--144.

\bibitem{Cella:1994px}
G.~Cella, G.~Curci, G.~Ricciardi, and A.~Vicere, {\it {The $b \to s \gamma$
  decay revisited}},  {\em Phys. Lett. B} {\bf 325} (1994) 227--234,
  [\href{http://arxiv.org/abs/hep-ph/9401254}{{\tt hep-ph/9401254}}].

\bibitem{Misiak:2015xwa}
M.~Misiak et~al., {\it {Updated NNLO QCD predictions for the weak radiative
  B-meson decays}},  {\em Phys. Rev. Lett.} {\bf 114} (2015) 221801,
  [\href{http://arxiv.org/abs/1503.01789}{{\tt arXiv:1503.01789}}].

\bibitem{Grinstein:1987pu}
B.~Grinstein and M.~B. Wise, {\it {Weak Radiative B Meson Decay as a Probe of
  the Higgs Sector}},  {\em Phys. Lett. B} {\bf 201} (1988) 274--278.

\bibitem{Hou:1987kf}
W.-S. Hou and R.~S. Willey, {\it {Effects of Charged Higgs Bosons on the
  Processes $b \to s \gamma$, $b \to s g^*$ and $b \to s \ell^+ \ell^-$}},
  {\em Phys. Lett. B} {\bf 202} (1988) 591--595.

\bibitem{Gabbiani:1996hi}
F.~Gabbiani, E.~Gabrielli, A.~Masiero, and L.~Silvestrini, {\it {A Complete
  analysis of FCNC and CP constraints in general SUSY extensions of the
  standard model}},  {\em Nucl. Phys. B} {\bf 477} (1996) 321--352,
  [\href{http://arxiv.org/abs/hep-ph/9604387}{{\tt hep-ph/9604387}}].

\bibitem{Everett:2001yy}
L.~L. Everett, G.~L. Kane, S.~Rigolin, L.-T. Wang, and T.~T. Wang, {\it
  {Alternative approach to $b \to s \gamma$ in the uMSSM}},  {\em JHEP} {\bf
  01} (2002) 022, [\href{http://arxiv.org/abs/hep-ph/0112126}{{\tt
  hep-ph/0112126}}].

\bibitem{Borzumati:2003rr}
F.~Borzumati, C.~Greub, and Y.~Yamada, {\it {Beyond leading order corrections
  to $\bar B \to X_s \gamma$ at large $tan(\beta)$: The Charged Higgs
  contribution}},  {\em Phys. Rev. D} {\bf 69} (2004) 055005,
  [\href{http://arxiv.org/abs/hep-ph/0311151}{{\tt hep-ph/0311151}}].

\bibitem{Buras:2003mk}
A.~J. Buras, A.~Poschenrieder, M.~Spranger, and A.~Weiler, {\it {The Impact of
  universal extra dimensions on $B \to X_s \gamma$, $B \to X_s$ gluon, $B \to
  X_s \mu^+ \mu^-$, $K_L \to \pi^0 e^+ e^-$ and $\epsilon^\prime/ \epsilon$}},
  {\em Nucl. Phys. B} {\bf 678} (2004) 455--490,
  [\href{http://arxiv.org/abs/hep-ph/0306158}{{\tt hep-ph/0306158}}].

\bibitem{Blanke:2006sb}
M.~Blanke, A.~J. Buras, A.~Poschenrieder, C.~Tarantino, S.~Uhlig, and
  A.~Weiler, {\it {Particle-Antiparticle Mixing, $\epsilon_K$, $\Delta
  \Gamma_q$, $A^q_{SL}$, $A_{CP} (B_d \to \psi K_S)$, $A_{CP} (B_s \to \psi
  \phi)$ and $B \to X_{s,d} \gamma$ in the Littlest Higgs Model with
  T-Parity}},  {\em JHEP} {\bf 12} (2006) 003,
  [\href{http://arxiv.org/abs/hep-ph/0605214}{{\tt hep-ph/0605214}}].

\bibitem{Misiak:2017bgg}
M.~Misiak and M.~Steinhauser, {\it {Weak radiative decays of the B meson and
  bounds on $M_{H^\pm }$ in the Two-Higgs-Doublet Model}},  {\em Eur. Phys. J.
  C} {\bf 77} (2017) 201, [\href{http://arxiv.org/abs/1702.04571}{{\tt
  arXiv:1702.04571}}].

\bibitem{Buras:2020xsm}
A.~J. Buras, {\em {Gauge Theory of Weak Decays}}.
\newblock Cambridge University Press, 6, 2020.

\bibitem{CLEO:1993nic}
{\bf CLEO} Collaboration, R.~Ammar et~al., {\it {Evidence for penguins: First
  observation of $B \to K^* (892) \gamma$}},  {\em Phys. Rev. Lett.} {\bf 71}
  (1993) 674--678.

\bibitem{HFLAV:2022pwe}
{\bf HFLAV} Collaboration, Y.~S. Amhis et~al., {\it {Averages of $b$-hadron,
  $c$-hadron, and $\tau$-lepton properties as of 2021}},
  \href{http://arxiv.org/abs/2206.07501}{{\tt arXiv:2206.07501}}.

\bibitem{ParticleDataGroup:2022pth}
{\bf Particle Data Group} Collaboration, R.~L. Workman et~al., {\it {Review of
  Particle Physics}},  {\em PTEP} {\bf 2022} (2022) 083C01.

\bibitem{LHCb:2019wwi}
{\bf LHCb} Collaboration, R.~Aaij et~al., {\it {First Observation of the
  Radiative Decay $\Lambda_{b}^{0} \to \Lambda \gamma$}},  {\em Phys. Rev.
  Lett.} {\bf 123} (2019) 031801, [\href{http://arxiv.org/abs/1904.06697}{{\tt
  arXiv:1904.06697}}].

\bibitem{LHCb:2021byf}
{\bf LHCb} Collaboration, R.~Aaij et~al., {\it {Measurement of the photon
  polarization in $\Lambda_{b}^{0}$ $\to$ $\Lambda$ $\gamma$ decays}},  {\em
  Phys. Rev. D} {\bf 105} (2022) L051104,
  [\href{http://arxiv.org/abs/2111.10194}{{\tt arXiv:2111.10194}}].

\bibitem{LHCb:2021hfz}
{\bf LHCb} Collaboration, R.~Aaij et~al., {\it {Search for the radiative
  $\Xi_b^- \to \Xi^- \gamma$ decay}},  {\em JHEP} {\bf 01} (2022) 069,
  [\href{http://arxiv.org/abs/2108.07678}{{\tt arXiv:2108.07678}}].

\bibitem{Colangelo:1993ux}
P.~Colangelo, C.~A. Dominguez, G.~Nardulli, and N.~Paver, {\it {Radiative $B
  \to K^* \gamma $ transition in QCD}},  {\em Phys. Lett. B} {\bf 317} (1993)
  183--189, [\href{http://arxiv.org/abs/hep-ph/9308264}{{\tt hep-ph/9308264}}].

\bibitem{Neubert:1993mb}
M.~Neubert, {\it {Heavy quark symmetry}},  {\em Phys. Rept.} {\bf 245} (1994)
  259--396, [\href{http://arxiv.org/abs/hep-ph/9306320}{{\tt hep-ph/9306320}}].

\bibitem{Kagan:1998bh}
A.~L. Kagan and M.~Neubert, {\it {Direct CP violation in $B \to X_s \gamma$
  decays as a signature of new physics}},  {\em Phys. Rev. D} {\bf 58} (1998)
  094012, [\href{http://arxiv.org/abs/hep-ph/9803368}{{\tt hep-ph/9803368}}].

\bibitem{Benzke:2010tq}
M.~Benzke, S.~J. Lee, M.~Neubert, and G.~Paz, {\it {Long-Distance Dominance of
  the CP Asymmetry in $B\to X_{s,d}+\gamma$ Decays}},  {\em Phys. Rev. Lett.}
  {\bf 106} (2011) 141801, [\href{http://arxiv.org/abs/1012.3167}{{\tt
  arXiv:1012.3167}}].

\bibitem{Catani:1997xc}
S.~Catani and B.~R. Webber, {\it {Infrared safe but infinite: Soft gluon
  divergences inside the physical region}},  {\em JHEP} {\bf 10} (1997) 005,
  [\href{http://arxiv.org/abs/hep-ph/9710333}{{\tt hep-ph/9710333}}].

\bibitem{DeFazio:1999ptt}
F.~De~Fazio and M.~Neubert, {\it {$B \to X_u \ell \bar \nu_\ell$ decay
  distributions to order $\alpha_s$}},  {\em JHEP} {\bf 06} (1999) 017,
  [\href{http://arxiv.org/abs/hep-ph/9905351}{{\tt hep-ph/9905351}}].

\bibitem{Neubert:1993ch}
M.~Neubert, {\it {QCD based interpretation of the lepton spectrum in inclusive
  $\bar B \to X_u$ lepton anti-neutrino decays}},  {\em Phys. Rev. D} {\bf 49}
  (1994) 3392--3398, [\href{http://arxiv.org/abs/hep-ph/9311325}{{\tt
  hep-ph/9311325}}].

\bibitem{Neubert:1993um}
M.~Neubert, {\it {Analysis of the photon spectrum in inclusive $B \to X_s
  \gamma$ decays}},  {\em Phys. Rev. D} {\bf 49} (1994) 4623--4633,
  [\href{http://arxiv.org/abs/hep-ph/9312311}{{\tt hep-ph/9312311}}].

\bibitem{Bigi:1993ex}
I.~I.~Y. Bigi, M.~A. Shifman, N.~G. Uraltsev, and A.~I. Vainshtein, {\it {On
  the motion of heavy quarks inside hadrons: Universal distributions and
  inclusive decays}},  {\em Int. J. Mod. Phys. A} {\bf 9} (1994) 2467--2504,
  [\href{http://arxiv.org/abs/hep-ph/9312359}{{\tt hep-ph/9312359}}].

\bibitem{Bernlochner:2020jlt}
{\bf SIMBA} Collaboration, F.~U. Bernlochner, H.~Lacker, Z.~Ligeti, I.~W.
  Stewart, F.~J. Tackmann, and K.~Tackmann, {\it {Precision Global
  Determination of the $B \to X_s \gamma$ Decay Rate}},  {\em Phys. Rev. Lett.}
  {\bf 127} (2021) 102001, [\href{http://arxiv.org/abs/2007.04320}{{\tt
  arXiv:2007.04320}}].

\bibitem{Benzke:2010js}
M.~Benzke, S.~J. Lee, M.~Neubert, and G.~Paz, {\it {Factorization at Subleading
  Power and Irreducible Uncertainties in $\bar B\to X_s \gamma$ Decay}},  {\em
  JHEP} {\bf 08} (2010) 099, [\href{http://arxiv.org/abs/1003.5012}{{\tt
  arXiv:1003.5012}}].

\bibitem{Donoghue:1995na}
J.~F. Donoghue and A.~A. Petrov, {\it {Is $B \to X_s \gamma$ equal to $b \to s
  \gamma$ ? Spectator contributions to rare inclusive B decays}},  {\em Phys.
  Rev. D} {\bf 53} (1996) 3664--3671,
  [\href{http://arxiv.org/abs/hep-ph/9510227}{{\tt hep-ph/9510227}}].

\bibitem{Lee:2006wn}
S.~J. Lee, M.~Neubert, and G.~Paz, {\it {Enhanced Non-local Power Corrections
  to the $\bar B \to X_s \gamma$ Decay Rate}},  {\em Phys. Rev. D} {\bf 75}
  (2007) 114005, [\href{http://arxiv.org/abs/hep-ph/0609224}{{\tt
  hep-ph/0609224}}].

\bibitem{Voloshin:1996gw}
M.~B. Voloshin, {\it {Large $O (m_c^{-2})$ nonperturbative correction to the
  inclusive rate of the decay $B \to X_s \gamma$}},  {\em Phys. Lett. B} {\bf
  397} (1997) 275--278, [\href{http://arxiv.org/abs/hep-ph/9612483}{{\tt
  hep-ph/9612483}}].

\bibitem{Ligeti:1997tc}
Z.~Ligeti, L.~Randall, and M.~B. Wise, {\it {Comment on nonperturbative effects
  in $\bar B \to X_s \gamma$}},  {\em Phys. Lett. B} {\bf 402} (1997) 178--182,
  [\href{http://arxiv.org/abs/hep-ph/9702322}{{\tt hep-ph/9702322}}].

\bibitem{Grant:1997ec}
A.~K. Grant, A.~G. Morgan, S.~Nussinov, and R.~D. Peccei, {\it {Comment on
  nonperturbative $O (1/m_c^2)$ corrections to $\Gamma (\bar B \to X_s
  \gamma)$}},  {\em Phys. Rev. D} {\bf 56} (1997) 3151--3154,
  [\href{http://arxiv.org/abs/hep-ph/9702380}{{\tt hep-ph/9702380}}].

\bibitem{Buchalla:1997ky}
G.~Buchalla, G.~Isidori, and S.~J. Rey, {\it {Corrections of order
  $\Lambda_{QCD}^2 / m_c^2$ to inclusive rare B decays}},  {\em Nucl. Phys. B}
  {\bf 511} (1998) 594--610, [\href{http://arxiv.org/abs/hep-ph/9705253}{{\tt
  hep-ph/9705253}}].

\bibitem{BUSKULIC1996437}
{\bf ALEPH} Collaboration, D.~Buskulic et~al., {\it {Measurement of the
  $\Lambda_b$ polarization in Z decays}},  {\em Phys. Lett. B} {\bf 365} (1996)
  437 -- 447.

\bibitem{Abbiendi:1998uz}
{\bf OPAL} Collaboration, G.~Abbiendi et~al., {\it {Measurement of the average
  polarization of b baryons in hadronic $Z^0$ decays}},  {\em Phys. Lett. B}
  {\bf 444} (1998) 539--554, [\href{http://arxiv.org/abs/hep-ex/9808006}{{\tt
  hep-ex/9808006}}].

\bibitem{2000205}
{\bf DELPHI} Collaboration, P.~Abreu et~al., {\it {$\Lambda_b$ polarization in
  $Z^0$ decays at LEP}},  {\em Phys. Lett. B} {\bf 474} (2000) 205 -- 222.

\bibitem{Colangelo:2020vhu}
P.~Colangelo, F.~De~Fazio, and F.~Loparco, {\it {Inclusive semileptonic
  $\Lambda_{b}$ decays in the Standard Model and beyond}},  {\em JHEP} {\bf 11}
  (2020) 032, [\href{http://arxiv.org/abs/2006.13759}{{\tt arXiv:2006.13759}}].
  [Erratum: JHEP 12, 098 (2022)].

\bibitem{Borzumati:1999qt}
F.~Borzumati, C.~Greub, T.~Hurth, and D.~Wyler, {\it {Gluino contribution to
  radiative B decays: Organization of QCD corrections and leading order
  results}},  {\em Phys. Rev. D} {\bf 62} (2000) 075005,
  [\href{http://arxiv.org/abs/hep-ph/9911245}{{\tt hep-ph/9911245}}].

\bibitem{Buras:1993xp}
A.~J. Buras, M.~Misiak, M.~Munz, and S.~Pokorski, {\it {Theoretical
  uncertainties and phenomenological aspects of $B \to X_s \gamma$ decay}},
  {\em Nucl. Phys. B} {\bf 424} (1994) 374--398,
  [\href{http://arxiv.org/abs/hep-ph/9311345}{{\tt hep-ph/9311345}}].

\bibitem{Misiak:2018cec}
M.~Misiak, {\it {Radiative Decays of the $B$ Meson: a Progress Report}},  {\em
  Acta Phys. Polon. B} {\bf 49} (2018) 1291--1300.

\bibitem{Chay:1990da}
J.~Chay, H.~Georgi, and B.~Grinstein, {\it {Lepton energy distributions in
  heavy meson decays from QCD}},  {\em Phys. Lett. B} {\bf 247} (1990)
  399--405.

\bibitem{Bigi:1993fe}
I.~I. Bigi, M.~A. Shifman, N.~Uraltsev, and A.~I. Vainshtein, {\it {QCD
  predictions for lepton spectra in inclusive heavy flavor decays}},  {\em
  Phys. Rev. Lett.} {\bf 71} (1993) 496--499,
  [\href{http://arxiv.org/abs/hep-ph/9304225}{{\tt hep-ph/9304225}}].

\bibitem{Falk:1993dh}
A.~F. Falk, M.~E. Luke, and M.~J. Savage, {\it {Nonperturbative contributions
  to the inclusive rare decays $B \to X_s \gamma$ and $B \to X_s \ell^+
  \ell^-$}},  {\em Phys. Rev. D} {\bf 49} (1994) 3367--3378,
  [\href{http://arxiv.org/abs/hep-ph/9308288}{{\tt hep-ph/9308288}}].

\bibitem{Kapustin:1995nr}
A.~Kapustin and Z.~Ligeti, {\it {Moments of the photon spectrum in the
  inclusive $B \to X_s \gamma$ decay}},  {\em Phys. Lett. B} {\bf 355} (1995)
  318--324, [\href{http://arxiv.org/abs/hep-ph/9506201}{{\tt hep-ph/9506201}}].

\bibitem{Bauer:1997fe}
C.~W. Bauer, {\it {Corrections to moments of the photon spectrum in the
  inclusive decay $B \to X_s \gamma$}},  {\em Phys. Rev. D} {\bf 57} (1998)
  5611--5619, [\href{http://arxiv.org/abs/hep-ph/9710513}{{\tt
  hep-ph/9710513}}]. [Erratum: Phys.Rev.D 60, 099907 (1999)].

\bibitem{Mannel:2018mqv}
T.~Mannel and K.~K. Vos, {\it {Reparametrization Invariance and Partial
  Re-Summations of the Heavy Quark Expansion}},  {\em JHEP} {\bf 06} (2018)
  115, [\href{http://arxiv.org/abs/1802.09409}{{\tt arXiv:1802.09409}}].

\bibitem{Gremm:1995nx}
M.~Gremm, F.~Kruger, and L.~M. Sehgal, {\it {Angular distribution and
  polarization of photons in the inclusive decay $\Lambda_b \to X_s \gamma$}},
  {\em Phys. Lett. B} {\bf 355} (1995) 579--583,
  [\href{http://arxiv.org/abs/hep-ph/9505354}{{\tt hep-ph/9505354}}].

\bibitem{Mannel:1997xy}
T.~Mannel and S.~Recksiegel, {\it {Flavor changing neutral current decays of
  heavy baryons: The Case $\Lambda_b \to \Lambda \gamma$}},  {\em J. Phys. G}
  {\bf 24} (1998) 979--990, [\href{http://arxiv.org/abs/hep-ph/9701399}{{\tt
  hep-ph/9701399}}].

\bibitem{Huang:1998ek}
C.-S. Huang and H.-G. Yan, {\it {Exclusive rare decays of heavy baryons to
  light baryons: $\Lambda(b) \to \Lambda \gamma$ and $\Lambda_b \to \Lambda
  \ell^+ \ell^-$}},  {\em Phys. Rev. D} {\bf 59} (1999) 114022,
  [\href{http://arxiv.org/abs/hep-ph/9811303}{{\tt hep-ph/9811303}}]. [Erratum:
  Phys.Rev.D 61, 039901 (2000)].

\bibitem{Hiller:2001zj}
G.~Hiller and A.~Kagan, {\it {Probing for new physics in polarized $\Lambda_b$
  decays at the $Z$}},  {\em Phys. Rev. D} {\bf 65} (2002) 074038,
  [\href{http://arxiv.org/abs/hep-ph/0108074}{{\tt hep-ph/0108074}}].

\bibitem{Colangelo:2007jy}
P.~Colangelo, F.~De~Fazio, R.~Ferrandes, and T.~N. Pham, {\it {FCNC $B_s$ and
  $\Lambda_b$ transitions: Standard model versus a single universal extra
  dimension scenario}},  {\em Phys. Rev. D} {\bf 77} (2008) 055019,
  [\href{http://arxiv.org/abs/0709.2817}{{\tt arXiv:0709.2817}}].

\bibitem{Oliver:2010im}
L.~Oliver, J.~C. Raynal, and R.~Sinha, {\it {Note on new interesting baryon
  channels to measure the photon polarization in $b \to s \gamma$}},  {\em
  Phys. Rev. D} {\bf 82} (2010) 117502,
  [\href{http://arxiv.org/abs/1007.3632}{{\tt arXiv:1007.3632}}].

\bibitem{Blake:2015tda}
T.~Blake, T.~Gershon, and G.~Hiller, {\it {Rare b hadron decays at the LHC}},
  {\em Ann. Rev. Nucl. Part. Sci.} {\bf 65} (2015) 113--143,
  [\href{http://arxiv.org/abs/1501.03309}{{\tt arXiv:1501.03309}}].

\bibitem{GarciaMartin:2019bxm}
L.~M. Garc\'\i{}a~Mart\'\i{}n, B.~Jashal, F.~Mart\'\i{}nez~Vidal, A.~Oyanguren,
  S.~Roy, R.~Sain, and R.~Sinha, {\it {Radiative $b$-baryon decays to measure
  the photon and $b$-baryon polarization}},  {\em Eur. Phys. J. C} {\bf 79}
  (2019) 634, [\href{http://arxiv.org/abs/1902.04870}{{\tt arXiv:1902.04870}}].

\bibitem{Paul:2016urs}
A.~Paul and D.~M. Straub, {\it {Constraints on new physics from radiative $B$
  decays}},  {\em JHEP} {\bf 04} (2017) 027,
  [\href{http://arxiv.org/abs/1608.02556}{{\tt arXiv:1608.02556}}].

\bibitem{Bauer:2003pi}
C.~W. Bauer and A.~V. Manohar, {\it {Shape function effects in $B \to X_s
  \gamma$ and $B \to X_u \ell \bar \nu$ decays}},  {\em Phys. Rev. D} {\bf 70}
  (2004) 034024, [\href{http://arxiv.org/abs/hep-ph/0312109}{{\tt
  hep-ph/0312109}}].

\bibitem{Bauer:2000yr}
C.~W. Bauer, S.~Fleming, D.~Pirjol, and I.~W. Stewart, {\it {An Effective field
  theory for collinear and soft gluons: Heavy to light decays}},  {\em Phys.
  Rev. D} {\bf 63} (2001) 114020,
  [\href{http://arxiv.org/abs/hep-ph/0011336}{{\tt hep-ph/0011336}}].

\bibitem{Bauer:2001ct}
C.~W. Bauer and I.~W. Stewart, {\it {Invariant operators in collinear effective
  theory}},  {\em Phys. Lett. B} {\bf 516} (2001) 134--142,
  [\href{http://arxiv.org/abs/hep-ph/0107001}{{\tt hep-ph/0107001}}].

\bibitem{Beneke:2002ph}
M.~Beneke, A.~P. Chapovsky, M.~Diehl, and T.~Feldmann, {\it {Soft collinear
  effective theory and heavy to light currents beyond leading power}},  {\em
  Nucl. Phys. B} {\bf 643} (2002) 431--476,
  [\href{http://arxiv.org/abs/hep-ph/0206152}{{\tt hep-ph/0206152}}].

\bibitem{Beneke:2002ni}
M.~Beneke and T.~Feldmann, {\it {Multipole expanded soft collinear effective
  theory with nonAbelian gauge symmetry}},  {\em Phys. Lett. B} {\bf 553}
  (2003) 267--276, [\href{http://arxiv.org/abs/hep-ph/0211358}{{\tt
  hep-ph/0211358}}].

\bibitem{Bosch:2004th}
S.~W. Bosch, B.~O. Lange, M.~Neubert, and G.~Paz, {\it {Factorization and shape
  function effects in inclusive B meson decays}},  {\em Nucl. Phys. B} {\bf
  699} (2004) 335--386, [\href{http://arxiv.org/abs/hep-ph/0402094}{{\tt
  hep-ph/0402094}}].

\bibitem{Gambino:2020jvv}
P.~Gambino et~al., {\it {Challenges in semileptonic $B$ decays}},  {\em Eur.
  Phys. J. C} {\bf 80} (2020), no.~10 966,
  [\href{http://arxiv.org/abs/2006.07287}{{\tt arXiv:2006.07287}}].

\bibitem{Mannel:1994pm}
T.~Mannel and M.~Neubert, {\it {Resummation of nonperturbative corrections to
  the lepton spectrum in inclusive $B \to X_q \ell \bar \nu_\ell$ decays}},
  {\em Phys. Rev. D} {\bf 50} (1994) 2037--2047,
  [\href{http://arxiv.org/abs/hep-ph/9402288}{{\tt hep-ph/9402288}}].

\bibitem{Benson:2004sg}
D.~Benson, I.~I. Bigi, and N.~Uraltsev, {\it {On the photon energy moments and
  their `bias' corrections in $B \to X_s + \gamma$}},  {\em Nucl. Phys. B} {\bf
  710} (2005) 371--401, [\href{http://arxiv.org/abs/hep-ph/0410080}{{\tt
  hep-ph/0410080}}].

\bibitem{Lange:2005yw}
B.~O. Lange, M.~Neubert, and G.~Paz, {\it {Theory of charmless inclusive B
  decays and the extraction of $V_{ub}$}},  {\em Phys. Rev. D} {\bf 72} (2005)
  073006, [\href{http://arxiv.org/abs/hep-ph/0504071}{{\tt hep-ph/0504071}}].

\bibitem{Andersen:2005mj}
J.~R. Andersen and E.~Gardi, {\it {Inclusive spectra in charmless semileptonic
  B decays by dressed gluon exponentiation}},  {\em JHEP} {\bf 01} (2006) 097,
  [\href{http://arxiv.org/abs/hep-ph/0509360}{{\tt hep-ph/0509360}}].

\bibitem{Gambino:2007rp}
P.~Gambino, P.~Giordano, G.~Ossola, and N.~Uraltsev, {\it {Inclusive
  semileptonic B decays and the determination of $|V_{ub}|$}},  {\em JHEP} {\bf
  10} (2007) 058, [\href{http://arxiv.org/abs/0707.2493}{{\tt
  arXiv:0707.2493}}].

\bibitem{Aglietti:2007ik}
U.~Aglietti, F.~Di~Lodovico, G.~Ferrera, and G.~Ricciardi, {\it {Inclusive
  measure of $|V_{ub}|$ with the analytic coupling model}},  {\em Eur. Phys. J.
  C} {\bf 59} (2009) 831--840, [\href{http://arxiv.org/abs/0711.0860}{{\tt
  arXiv:0711.0860}}].

\bibitem{MannelMainz}
T.~Mannel, {\it {Inclusive Semi-Leptonic B Decays}},  in {\em {Pushing the
  Limits of the Theoretical Physics, Mainz, 08-12 May 2023,
  indico.mitp.uni-mainz.de/event/341}}.

\bibitem{Ligeti:2008ac}
Z.~Ligeti, I.~W. Stewart, and F.~J. Tackmann, {\it {Treating the b quark
  distribution function with reliable uncertainties}},  {\em Phys. Rev. D} {\bf
  78} (2008) 114014, [\href{http://arxiv.org/abs/0807.1926}{{\tt
  arXiv:0807.1926}}].

\bibitem{Altarelli:1982kh}
G.~Altarelli, N.~Cabibbo, G.~Corbo, L.~Maiani, and G.~Martinelli, {\it
  {Leptonic Decay of Heavy Flavors: A Theoretical Update}},  {\em Nucl. Phys.
  B} {\bf 208} (1982) 365--380.

\bibitem{Dassinger:2006md}
B.~M. Dassinger, T.~Mannel, and S.~Turczyk, {\it {Inclusive semi-leptonic B
  decays to order $1 / m_b^4$}},  {\em JHEP} {\bf 03} (2007) 087,
  [\href{http://arxiv.org/abs/hep-ph/0611168}{{\tt hep-ph/0611168}}].

\bibitem{Mannel:2010wj}
T.~Mannel, S.~Turczyk, and N.~Uraltsev, {\it {Higher Order Power Corrections in
  Inclusive B Decays}},  {\em JHEP} {\bf 11} (2010) 109,
  [\href{http://arxiv.org/abs/1009.4622}{{\tt arXiv:1009.4622}}].

\end{thebibliography}\endgroup
\end{document}